\documentclass[12pt]{article}
\usepackage{amsfonts}
\usepackage[dvips]{graphicx}
\usepackage{latexsym,amsmath,amssymb,graphics,stmaryrd}
\usepackage{cite}
\usepackage{array}
\usepackage{subfigure}
\usepackage{multirow}
\usepackage{epsfig}
\usepackage[dvips]{graphicx}
\usepackage[dvips]{color}
\usepackage{pstricks}
\usepackage{pst-node}
\usepackage{pst-plot}
\usepackage{dsfont}
\usepackage{amsthm}
\usepackage{graphics}
\usepackage{subfigure}
\usepackage{fancyhdr}
\usepackage[dvips]{color}

\usepackage{feynmp}
\usepackage{Dalgebra}
\usepackage[ps2pdf=true,colorlinks]{hyperref}
\fancypagestyle{plain}{
\fancyhf{}
\newcommand{\fpage}{\iffloatpage{}{\thepage}}
\fancyfoot[C]{\fpage}

}

\newcommand{\be}{\begin{eqnarray}}
\newcommand{\ee}{\end{eqnarray}}

\newcommand{\la}{\lambda}

\newcommand{\NN}{\mathcal{N}}
\newcommand{\DD}{\mathcal{D}}

\newcommand{\OO}{\mathcal{O}}

\newcommand{\al}{\alpha}
\newcommand{\hh}{\smash{\widehat h}}
\newcommand{\sfrac}[2]{\textstyle{\frac{#1}{#2}}}

\newcommand{\eps}{\epsilon}
\newcommand{\veps}{\varepsilon}

\newlength{\neglength}
\newlength{\diameter}
    
  \newcommand{\nn}{\nonumber}

\DeclareMathOperator{\Tr}{Tr}

\DeclareMathOperator{\diag}{diag}
\numberwithin{equation}{section}
\newlength{\eqoff}

\newlength{\unit}
\psset{xunit=\unit,yunit=\unit,runit=\unit}
\newlength{\linew}
\psset{linewidth=\linew}

\newcommand{\nvml}[3][1]{%
\fmfcmd{%
begingroup;
save a, vp, tvp, nvp, tv, nv, ip, ts, tt, is, it, n, m, scale, t, r, s, ttpr, tnpr, ep, mm;
path lcirc;
pair vp[][], tvp[][], tv[][], nvp[][], nv[][], ip[][], ts[], is[], tt[], it[], ep[], mid;
n := #2;
m:=3;
for i=1 upto n:
for j=1 upto m:
a[i][j] := arctime ((j-1)/(m-1)*arclength pm[i]) of pm[i];
vp[i][j] := point a[i][j] of pm[i];
tvp[i][j] := unitvector direction a[i][j] of pm[i];
nvp[i][j] := tvp[i][j] rotated -90;
endfor;
endfor;
if(vp[1][1]=vp[n][m]):
vp[n+1][1] := vp[1][1];
tvp[0][m] := tvp[n][m];
nvp[0][m] := nvp[n][m];
tvp[n+1][1] :=tvp[1][1];
nvp[n+1][1] :=nvp[1][1];
else:
vp[n+1][1] := vp[n][m];
tvp[0][m] := (0,0);
nvp[0][m] := (0,0);
tvp[n+1][1] :=tvp[n][m];
nvp[n+1][1] :=nvp[n][m];
fi;
s := 1;
for i=1 upto n:
for j=1 upto m:
if (j=1):
tv[i][1] := (tvp[i-1][m]+tvp[i][1]);
nv[i][1] := (nvp[i-1][m]+nvp[i][1]);
if (not(tv[i][1]=(0,0))):
tv[i][1] := unitvector tv[i][1];
fi;
if (not(nv[i][1]=(0,0))):
nv[i][1] := unitvector nv[i][1];
fi;
ttpr := tvp[i][1] dotprod tvp[i-1][m];
tnpr := tvp[i][1] dotprod nvp[i-1][m];
elseif (j=m):
tv[i][m] := (tvp[i][m]+tvp[i+1][1]);
nv[i][m] := (nvp[i][m]+nvp[i+1][1]);
if (not(tv[i][m]=(0,0))):
tv[i][m] := unitvector tv[i][m];
fi;
if (not(nv[i][m]=(0,0))):
nv[i][m] := unitvector nv[i][m];
fi;
ttpr := tvp[i][m] dotprod tvp[i+1][1];
tnpr := -tvp[i][m] dotprod nvp[i+1][1];
else:
nv[i][j] :=nvp[i][j];
tv[i][j] :=tvp[i][j];
fi;
scale := 25;
if ((j=1) or (j=m)):
 if ((tnpr<=0) and not((tv[i][j]=(0,0)) or (nv[i][j]=(0,0)))):
  ip[i][j] := vp[i][j] shifted(0.15*scale*nvp[i][j]);
  ts[s] := tvp[i][j];
  is[s] := ip[i][j];
  s:=s+1;
 else:
  if ((j=1) and (ttpr>0)):
  fi;
 fi;
else:
 ip[i][j] := vp[i][j] shifted(0.15*scale*nv[i][j]);
 ts[s] := tv[i][j];
 is[s] := ip[i][j];
 s:=s+1;
fi;
endfor;
endfor;
if(vp[1][1]=vp[n][m]):
ts[s] := ts[1];
is[s] := is[1];
else:
tv[n+1][1] := unitvector (tvp[n][m]+tvp[n+1][1]);
nv[n+1][1] := unitvector (nvp[n][m]+nvp[n+1][1]);
ip[n+1][1] := vp[n+1][1] shifted(0.15*scale*nv[n+1][1]);
ts[s] := tv[n+1][1];
is[s] := ip[n+1][1];
fi;
t=#1;
lcirc:=is[1];
for k=2 upto s:
lcirc := lcirc{ts[k-1]}..tension t..{ts[k]}is[k];
endfor;
mm := arctime (0.5* arclength lcirc) of lcirc;
if(vp[1][1]=vp[n][m]):
ep1 := point arctime (0* arclength lcirc) of lcirc of lcirc;
ep2 := point mm of lcirc;
mid := 1/2[ep1,ep2];
else:
ep1 := point mm of lcirc;
ep2 :=unitvector direction mm of lcirc rotated -90;
mid:= ep1 shifted(0.2*scale*ep2);
fi;
draw(lcirc) withpen pencircle scaled 0.25;
drawarrow(subpath(mm*0.8,mm*1.1) of lcirc) withpen pencircle scaled 0.25;
endgroup;
}
\fmfiv{label=#3,l.dist=0}{mid}
}

\newcommand{\cvert}[7][]{%
\settoheight{\eqoff}{$\times$}%
\setlength{\eqoff}{0.5\eqoff}%
\addtolength{\eqoff}{-8\unitlength}%
\raisebox{\eqoff}{%
\fmfframe(1,2)(1,2){%
\begin{fmfchar*}(12,12)
\fmfright{v3,v2}
\fmfpoly{phantom}{v1,v3,v2}
\fmf{#2}{v1,vc1}
\fmf{#3}{vc1,v2}
\fmf{#4}{vc1,v3}
\fmffreeze
\fmfposition
\fmfipath{p[]}
\fmfiset{p1}{vpath(__v1,__vc1)}
\fmfiset{p2}{vpath(__vc1,__v2)}
\fmfiset{p3}{vpath(__vc1,__v3)}
{#1}
\fmfis{phantom,ptext.clen=6,ptext.hout=3,ptext.oout=12,ptext.out=#5,ptext.sep=;}{p1}
\fmfis{phantom,ptext.clen=6,ptext.hin=3,ptext.oin=14,ptext.in=#6,ptext.sep=;}{p2}
\fmfis{phantom,ptext.side=right,ptext.clen=6,ptext.hin=-3,ptext.oin=14,ptext.in=#7,ptext.sep=;}{p3}
\end{fmfchar*}}}
}

\newcommand{\qvert}[8]{%
\settoheight{\eqoff}{$\times$}%
\setlength{\eqoff}{0.5\eqoff}%
\addtolength{\eqoff}{-8\unitlength}%
\raisebox{\eqoff}{%
\fmfframe(1,2)(1,2){%
\begin{fmfchar*}(12,12)
\fmfleft{v2,v1}
\fmfright{v3,v4}
\fmfforce{(0,0)}{v1}
\fmfforce{(0,h)}{v2}
\fmfforce{(w,h)}{v3}
\fmfforce{(w,0)}{v4}
\fmf{#1}{v1,vc1}
\fmf{#2}{v2,vc1}
\fmf{#3}{vc1,v3}
\fmf{#4}{vc1,v4}
\fmffreeze
\fmfposition
\fmfipath{p[]}
\fmfiset{p1}{vpath(__v1,__vc1)}
\fmfiset{p2}{vpath(__v2,__vc1)}
\fmfiset{p3}{vpath(__vc1,__v3)}
\fmfiset{p4}{vpath(__vc1,__v4)}
\fmfis{phantom,ptext.clen=6,ptext.hout=3,ptext.oout=12,ptext.out=#6,ptext.sep=;}{p2}
\fmfis{phantom,ptext.clen=6,ptext.hout=3,ptext.oout=12,ptext.out=#5,ptext.sep=;}{p1}
\fmfis{phantom,ptext.side=right,ptext.clen=6,ptext.hin=-3,ptext.oin=12,ptext.in=#8,ptext.sep=;}{p4}
\fmfis{phantom,ptext.clen=6,ptext.hin=3,ptext.oin=12,ptext.in=#7,ptext.sep=;}{p3}
\end{fmfchar*}}}
}

\usepackage{rotating}
\newlength{\arlength}

\DeclareMathOperator{\D}{D}

\DeclareMathOperator{\barD}{\vphantom{\D}\smash[t]{\bar{\mathrm{D}}}}
\newcommand{\YM}{{\text{\tiny YM}}}

\newcommand{\svertex}[3][0.5]{%
\fmfiequ{#2}{point #1*length(#3) of #3}
}
\newcommand{\dvertex}[3]{%
\fmfiequ{#1}{point length(#3)/3 of #3}
\fmfiequ{#2}{point 2length(#3)/3 of #3}
}
\newcommand{\fvertex}[3]{%
\fmfiequ{#1}{point #2length(#3) of #3}
}

\newcommand{\chione}[1][black]{%
\fmftop{v1}
\fmfbottom{v3}
\fmfforce{(0.125w,h)}{v1}
\fmfforce{(0.125w,0)}{v3}
\fmffixed{(0.25w,0)}{v1,v2}
\fmffixed{(0.25w,0)}{v3,v4}
\fmf{plain,tension=0.5,right=0.25,fore=#1}{v1,vc1}
\fmf{plain,tension=0.5,left=0.25,fore=#1}{v2,vc1}
\fmf{plain,tension=1.25,fore=#1}{vc1,vc2}
\fmf{plain,tension=0.5,left=0.125,fore=#1}{vc3,vc2}
\fmf{plain,tension=0.5,left=0.25,fore=#1}{v3,vc2}
\fmf{plain,tension=0.5,right=0.25,fore=#1}{v4,vc2}
\fmf{plain,tension=0.5,right=0,width=1mm,fore=#1}{v3,v4}
\fmfposition
\fmfipath{p[]}
\fmfipair{vd[],vm[],vu[]}
\fmfiset{p1}{vpath(__v1,__vc1)}
\fmfiset{p2}{vpath(__v2,__vc1)}
\fmfiset{p3}{vpath(__vc1,__vc2)}
\fmfiset{p4}{vpath(__v3,__vc2)}
\fmfiset{p5}{vpath(__v4,__vc2)}
\svertex{vm1}{p1}
\dvertex{vu1}{vd1}{p1}
\svertex{vm2}{p2}
\dvertex{vu2}{vd2}{p2}
\svertex{vm3}{p3}
\dvertex{vu3}{vd3}{p3}
\svertex{vm4}{p4}
\dvertex{vd4}{vu4}{p4}
\svertex{vm5}{p5}
\dvertex{vd5}{vu5}{p5}
}

\newcommand{\chionen}[1][black]{%
\fmftop{v1}
\fmfbottom{v3}
\fmfforce{(0.125w,h)}{v1}
\fmfforce{(0.125w,0)}{v3}
\fmffixed{(0.25w,0)}{v1,v2}
\fmffixed{(0.25w,0)}{v3,v4}
\fmf{plain,tension=0.5,right=0.25,fore=#1}{v1,vc1}
\fmf{plain,tension=0.5,left=0.25,fore=#1}{v2,vc1}
\fmf{plain,tension=1.25,fore=#1}{vc1,vc2}
\fmf{plain,tension=0.5,left=0.125,fore=#1}{vc3,vc2}
\fmf{plain,tension=0.5,left=0.25,fore=#1}{v3,vc2}
\fmf{plain,tension=0.5,right=0.25,fore=#1}{v4,vc2}
\fmfposition
\fmfipath{p[]}
\fmfipair{vd[],vm[],vu[]}
\fmfiset{p1}{vpath(__v1,__vc1)}
\fmfiset{p2}{vpath(__v2,__vc1)}
\fmfiset{p3}{vpath(__vc1,__vc2)}
\fmfiset{p4}{vpath(__v3,__vc2)}
\fmfiset{p5}{vpath(__v4,__vc2)}
\svertex{vm1}{p1}
\dvertex{vu1}{vd1}{p1}
\svertex{vm2}{p2}
\dvertex{vu2}{vd2}{p2}
\svertex{vm3}{p3}
\dvertex{vu3}{vd3}{p3}
\svertex{vm4}{p4}
\dvertex{vd4}{vu4}{p4}
\svertex{vm5}{p5}
\dvertex{vd5}{vu5}{p5}
}

\newcommand{\chionefour}[1][black]{%
\fmftop{v1}
\fmfbottom{v3}
\fmfforce{(0.125w,h)}{v1}
\fmfforce{(0.125w,0)}{v3}
\fmffixed{(0.25w,0)}{v1,v2}
\fmffixed{(0.25w,0)}{v3,v4}
\fmf{plain,tension=0.5,right=0.25,fore=#1}{v1,vc1}
\fmf{plain,tension=0.5,left=0.25,fore=#1}{v2,vc1}
\fmf{plain,tension=1.25,fore=#1}{vc1,vc2}
\fmf{plain,tension=0.5,left=0.125,fore=#1}{vc3,vc2}
\fmf{plain,tension=0.5,left=0.25,fore=#1}{v3,vc2}
\fmf{plain,tension=0.5,right=0.25,fore=#1}{v4,vc2}
\fmf{plain,tension=0.5,right=0,width=1mm,fore=#1}{v3,v4}
\fmffixed{(0,.11h)}{vm1,vc2}
\fmffixed{(0,.22h)}{vm2,vc2}
\fmffreeze
\fmf{plain}{vm1,vm2}
\fmfposition
\fmfipath{p[]}
\fmfipair{vd[],vm[],vu[],vt[]}
\fmfiset{p1}{vpath(__v1,__vc1)}
\fmfiset{p2}{vpath(__v2,__vc1)}
\fmfiset{p3}{vpath(__vc1,__vc2)}
\fmfiset{p4}{vpath(__v3,__vc2)}
\fmfiset{p5}{vpath(__v4,__vc2)}
\svertex{vm1}{p1}
\dvertex{vu1}{vd1}{p1}
\svertex{vm2}{p2}
\dvertex{vu2}{vd2}{p2}
\svertex{vm3}{p3}
\dvertex{vu3}{vd3}{p3}
\svertex{vm4}{p4}
\dvertex{vd4}{vu4}{p4}
\svertex{vm5}{p5}
\dvertex{vd5}{vu5}{p5}
\fvertex{vt1}{5/6}{p5}
\fvertex{vt2}{1/2}{p5}
\fvertex{vt3}{1/6}{p5}
\fmfi{plain}{vd4--vd5}
\fmfi{plain}{vu4--vu5}
}

\newcommand{\nomix}[1][black]{%
\fmftop{v1}
\fmfbottom{v5}
\fmfforce{(0.125w,h)}{v1}
\fmfforce{(0.125w,0)}{v5}
\fmffixed{(0.25w,0)}{v1,v2}
\fmffixed{(0.25w,0)}{v2,v3}
\fmffixed{(0.25w,0)}{v3,v4}
\fmffixed{(0.25w,0)}{v5,v6}
\fmffixed{(0.25w,0)}{v6,v7}
\fmffixed{(0.25w,0)}{v7,v8}
\fmf{plain,tension=1}{v1,vc1}
\fmf{plain,tension=1}{v5,vc1}
\fmf{plain,tension=1}{v2,vc2}
\fmf{plain,tension=0.5,left=0.5}{vc2,vc3}
\fmf{plain,tension=0.5,left=0.5}{vc3,vc2}
\fmf{plain,tension=1}{v6,vc3}
\fmf{plain,tension=1}{v4,vc4}
\fmf{plain,tension=1}{v8,vc4}
\fmf{plain,tension=1}{v3,vc5}
\fmf{plain,tension=0.5,left=0.5}{vc5,vc6}
\fmf{plain,tension=0.5,left=0.5}{vc6,vc5}
\fmf{plain,tension=1}{v7,vc6}
\fmf{plain,tension=0.5,right=0,width=1mm,fore=#1}{v5,v8}
\fmfposition
\fmfipath{pl[],pr[]}
\fmfipair{vld[],vlm[],vlu[],vrd[],vrm[],vru[]}
\fmfiset{pl1}{vpath(__vc2,__vc3)}
\fmfiset{pl2}{vpath(__vc3,__vc2)}
\fmfiset{pl3}{vpath(__v6,__vc3)}
\fmfiset{pl5}{vpath(__v5,__vc1)}
\fmfiset{pl4}{vpath(__v1,__vc1)}
\fmfiset{pr1}{vpath(__vc5,__vc6)}
\fmfiset{pr2}{vpath(__vc6,__vc5)}
\fmfiset{pr3}{vpath(__v7,__vc6)}
\fmfiset{pr5}{vpath(__v8,__vc4)}
\fmfiset{pr4}{vpath(__v4,__vc4)}
\svertex{vlm1}{pl1}
\dvertex{vlu1}{vld1}{pl1}
\svertex{vlm2}{pl2}
\dvertex{vlu2}{vld2}{pl2}
\svertex{vlm3}{pl3}
\dvertex{vlu3}{vld3}{pl3}
\svertex{vlm4}{pl4}
\dvertex{vld4}{vlu4}{pl4}
\svertex{vlm5}{pl5}
\dvertex{vld5}{vlu5}{pl5}
\svertex{vrm1}{pr1}
\dvertex{vru1}{vrd1}{pr1}
\svertex{vrm2}{pr2}
\dvertex{vru2}{vrd2}{pr2}
\svertex{vrm3}{pr3}
\dvertex{vru3}{vrd3}{pr3}
\svertex{vrm4}{pr4}
\dvertex{vrd4}{vru4}{pr4}
\svertex{vrm5}{pr5}
\dvertex{vrd5}{vru5}{pr5}

}

\newcommand{\chioneg}[1][black]{%
\fmftop{v1}
\fmfbottom{v4}
\fmfforce{(0.125w,h)}{v1}
\fmfforce{(0.125w,0)}{v4}
\fmffixed{(0.25w,0)}{v1,v2}
\fmffixed{(0.25w,0)}{v2,v3}
\fmffixed{(0.25w,0)}{v4,v5}
\fmffixed{(0.25w,0)}{v5,v6}
\fmf{plain,tension=0.5,right=0.25,fore=#1}{v1,vc1}
\fmf{plain,tension=0.5,left=0.25,fore=#1}{v2,vc1}
  \fmf{plain,tension=1.25,fore=#1}{vc1,vc2}
\fmf{plain,tension=0.5,left=0.25,fore=#1}{v4,vc2}
\fmf{plain,tension=0.5,right=0.25,fore=#1}{v5,vc2}
\fmf{plain,fore=#1}{v3,v6}
\fmf{plain,tension=0.5,right=0,width=1mm,fore=#1}{v4,v6}
\fmfposition
\fmfipath{p[],pg}
\fmfipair{vd[],vm[],vu[],vgd[],vgm[],vgu[],vg[]}
\fmfiset{p1}{vpath(__v1,__vc1)}
\fmfiset{p2}{vpath(__v2,__vc1)}
\fmfiset{p3}{vpath(__vc1,__vc2)}
\fmfiset{p5}{vpath(__v5,__vc2)}
\fmfiset{p4}{vpath(__v4,__vc2)}
\fmfiset{pg}{vpath(__v3,__v6)}
\svertex{vm1}{p1}
\dvertex{vu1}{vd1}{p1}
\svertex{vm2}{p2}
\dvertex{vu2}{vd2}{p2}
\svertex{vm3}{p3}
\dvertex{vu3}{vd3}{p3}
\svertex{vm4}{p4}
\dvertex{vd4}{vu4}{p4}
\svertex{vm5}{p5}
\dvertex{vd5}{vu5}{p5}
\vvertex{vgu2}{vu2}{pg}
\vvertex{vgm2}{vm2}{pg}
\vvertex{vgd2}{vd2}{pg}
\vvertex{vgu3}{vu3}{pg}
\vvertex{vgm3}{vm3}{pg}
\vvertex{vgd3}{vd3}{pg}
\vvertex{vgu5}{vu5}{pg}
\vvertex{vgm5}{vm5}{pg}
\vvertex{vgd5}{vd5}{pg}
\vvertex{vg1}{vloc(__vc1)}{pg}
\vvertex{vg2}{vloc(__vc2)}{pg}
}

\newcommand{\chionefourg}[1][black]{%
\fmftop{v1}
\fmfbottom{v4}
\fmfforce{(0.125w,h)}{v1}
\fmfforce{(0.125w,0)}{v4}
\fmffixed{(0.25w,0)}{v1,v2}
\fmffixed{(0.25w,0)}{v2,v3}
\fmffixed{(0.25w,0)}{v4,v5}
\fmffixed{(0.25w,0)}{v5,v6}
\fmf{plain,tension=0.5,right=0.25,fore=#1}{v1,vc1}
\fmf{plain,tension=0.5,left=0.25,fore=#1}{v2,vc1}
  \fmf{plain,tension=1.25,fore=#1}{vc1,vc2}
\fmf{plain,tension=0.5,left=0.25,fore=#1}{v4,vc2}
\fmf{plain,tension=0.5,right=0.25,fore=#1}{v5,vc2}
\fmf{plain,fore=#1}{v3,v6}
\fmf{plain,tension=0.5,right=0,width=1mm,fore=#1}{v4,v6}
\fmffixed{(0,.11h)}{vm1,vc2}
\fmffixed{(0,.22h)}{vm2,vc2}
\fmffreeze
\fmf{plain}{vm1,vm2}
\fmfposition
\fmfipath{p[],pg}
\fmfipair{vd[],vm[],vu[],vgd[],vgm[],vgu[],vg[],vf[],vgf[]}
\fmfiset{p1}{vpath(__v1,__vc1)}
\fmfiset{p2}{vpath(__v2,__vc1)}
\fmfiset{p3}{vpath(__vc1,__vc2)}
\fmfiset{p5}{vpath(__v5,__vc2)}
\fmfiset{p4}{vpath(__v4,__vc2)}
\fmfiset{pg}{vpath(__v3,__v6)}
\svertex{vm1}{p1}
\dvertex{vu1}{vd1}{p1}
\svertex{vm2}{p2}
\dvertex{vu2}{vd2}{p2}
\svertex{vm3}{p3}
\dvertex{vu3}{vd3}{p3}
\svertex{vm4}{p4}
\dvertex{vd4}{vu4}{p4}
\svertex{vm5}{p5}
\dvertex{vd5}{vu5}{p5}
\fvertex{vf51}{1/6}{p5}
\fvertex{vf55}{5/6}{p5}
\vvertex{vgu2}{vu2}{pg}
\vvertex{vgm2}{vm2}{pg}
\vvertex{vgd2}{vd2}{pg}
\vvertex{vgu3}{vu3}{pg}
\vvertex{vgm3}{vm3}{pg}
\vvertex{vgd3}{vd3}{pg}
\vvertex{vgu5}{vu5}{pg}
\vvertex{vgm5}{vm5}{pg}
\vvertex{vgf51}{vf51}{pg}
\vvertex{vgf55}{vf55}{pg}
\vvertex{vgd5}{vd5}{pg}
\vvertex{vg1}{vloc(__vc1)}{pg}
\vvertex{vg2}{vloc(__vc2)}{pg}
\fmfi{plain}{vd4--vd5}
\fmfi{plain}{vu4--vu5}
}

\newcommand{\vvertex}[3]{%
\fmfipath{px}
\fmfiequ{px}{(0,ypart(#2))..(100,ypart(#2))}
\fmfiequ{#1}{point xpart(#3 intersectiontimes px) of #3}
}

\newcommand{\chionetwon}[1][black]{%
\fmftop{v1}
\fmfbottom{v4}
\fmfforce{(0.125w,h)}{v1}
\fmfforce{(0.125w,0)}{v4}
\fmffixed{(0.25w,0)}{v1,v2}
\fmffixed{(0.25w,0)}{v2,v3}
\fmffixed{(0.25w,0)}{v4,v5}
\fmffixed{(0.25w,0)}{v5,v6}
\fmffixed{(0,whatever)}{vc1,vc3}
\fmffixed{(0,whatever)}{vc2,vc4}
\fmf{plain,tension=1.0,right=0.25}{v1,vc1}
\fmf{plain,tension=1.0,left=0.25}{v2,vc1}
\fmf{phantom,tension=0.3,right=0.25}{v2,vc2}
\fmf{plain,tension=0.3,left=0.25}{v3,vc2}
\fmf{plain,tension=0.3,left=0.25}{v4,vc3}
\fmf{phantom,tension=0.3,right=0.25}{v5,vc3}
\fmf{plain,tension=1.0,left=0.25}{v5,vc4}
\fmf{plain,tension=1.0,right=0.25}{v6,vc4}
\fmf{plain,tension=1.75,left=0}{vc1,vc3}
\fmf{plain,tension=1.75,left=0}{vc2,vc4}
\fmffreeze
\fmf{plain,tension=1,right=0.0}{vc2,vc3}
\fmffreeze
\fmfposition
\fmfipath{p[]}
\fmfipair{vd[],vm[],vu[]}
\fmfiset{p1}{vpath(__v1,__vc1)}
\fmfiset{p2}{vpath(__v2,__vc1)}
\fmfiset{p6}{vpath(__v3,__vc2)}
\fmfiset{p4}{vpath(__v4,__vc3)}
\fmfiset{p8}{vpath(__v5,__vc4)}
\fmfiset{p9}{vpath(__v6,__vc4)}
\fmfiset{p3}{vpath(__vc1,__vc3)}
\fmfiset{p7}{vpath(__vc2,__vc4)}
\fmfiset{p5}{vpath(__vc2,__vc3)}
\svertex{vm1}{p1}
\svertex{vm2}{p2}
\svertex{vm3}{p3}
\svertex{vm4}{p4}
\svertex{vm5}{p5}
\svertex{vm6}{p6}
\svertex{vm7}{p7}
\svertex{vm8}{p8}
\svertex{vm9}{p9}
}

\newcommand{\chitwoonen}[1][black]{%
\fmftop{v1}
\fmfbottom{v4}
\fmfforce{(0.125w,h)}{v1}
\fmfforce{(0.125w,0)}{v4}
\fmffixed{(0.25w,0)}{v1,v2}
\fmffixed{(0.25w,0)}{v2,v3}
\fmffixed{(0.25w,0)}{v4,v5}
\fmffixed{(0.25w,0)}{v5,v6}
\fmffixed{(0,whatever)}{vc1,vc3}
\fmffixed{(0,whatever)}{vc2,vc4}
\fmf{plain,tension=1.0,left=0.25}{v3,vc1}
\fmf{plain,tension=1.0,right=0.25}{v2,vc1}
\fmf{phantom,tension=0.3,left=0.25}{v2,vc2}
\fmf{plain,tension=0.3,right=0.25}{v1,vc2}
\fmf{plain,tension=0.3,right=0.25}{v6,vc3}
\fmf{phantom,tension=0.3,left=0.25}{v5,vc3}
\fmf{plain,tension=1.0,right=0.25}{v5,vc4}
\fmf{plain,tension=1.0,left=0.25}{v4,vc4}
\fmf{plain,tension=1.75,left=0}{vc1,vc3}
\fmf{plain,tension=1.75,left=0}{vc2,vc4}
\fmffreeze
\fmf{plain,tension=1,left=0.0}{vc2,vc3}
\fmffreeze
\fmfposition
\fmfipath{p[]}
\fmfipair{vd[],vm[],vu[]}
\fmfiset{p1}{vpath(__v3,__vc1)}
\fmfiset{p2}{vpath(__v2,__vc1)}
\fmfiset{p6}{vpath(__v1,__vc2)}
\fmfiset{p4}{vpath(__v6,__vc3)}
\fmfiset{p8}{vpath(__v5,__vc4)}
\fmfiset{p9}{vpath(__v4,__vc4)}
\fmfiset{p3}{vpath(__vc1,__vc3)}
\fmfiset{p7}{vpath(__vc2,__vc4)}
\fmfiset{p5}{vpath(__vc2,__vc3)}
\svertex{vm1}{p1}
\svertex{vm2}{p2}
\svertex{vm3}{p3}
\svertex{vm4}{p4}
\svertex{vm5}{p5}
\svertex{vm6}{p6}
\svertex{vm7}{p7}
\svertex{vm8}{p8}
\svertex{vm9}{p9}
}

\newcommand{\chioneonen}[1][black]{%
\fmftop{v1}
\fmfbottom{v3}
\fmfforce{(0.333w,h)}{v1}
\fmfforce{(0.333w,0)}{v3}
\fmffixed{(0.333w,0)}{v1,v2}
\fmffixed{(0.333w,0)}{v3,v4}
\fmffixed{(0,whatever)}{vc1,vc2}
\fmffixed{(0,whatever)}{vc2,vc3}
\fmffixed{(0,whatever)}{vc3,vc4}
\fmf{plain,tension=0.5,right=0.25}{v1,vc1}
\fmf{plain,tension=0.5,left=0.25}{v2,vc1}
\fmf{plain,tension=0.5,left=0.25}{v3,vc4}
\fmf{plain,tension=0.5,right=0.25}{v4,vc4}
\fmf{plain,tension=1.0,left=0.0}{vc1,vc2}
\fmf{plain,tension=0.2,right=0.7}{vc2,vc3}
\fmf{plain,tension=0.2,left=0.7}{vc2,vc3}
\fmf{plain,tension=1.0,left=0.0}{vc4,vc3}
\fmffreeze
}

\newcommand{\chionetwoone}[1][black]{%
\fmftop{v1}
\fmfbottom{v4}
\fmfforce{(0.125w,h)}{v1}
\fmfforce{(0.125w,0)}{v4}
\fmffixed{(0.25w,0)}{v1,v2}
\fmffixed{(0.25w,0)}{v2,v3}
\fmffixed{(0.25w,0)}{v4,v5}
\fmffixed{(0.25w,0)}{v5,v6}
\fmffixed{(0,whatever)}{vc1,vc3}
\fmffixed{(0,whatever)}{vb2,vb4}
\fmffixed{(0,whatever)}{vc1,vb1}
\fmffixed{(0,whatever)}{vc1,vb3}
\fmffixed{(whatever,0)}{vb1,vb2}
\fmffixed{(whatever,0)}{vb3,vb4}
\fmf{plain,tension=0.5,right=0.25}{v1,vc1}
\fmf{plain,tension=0.5,left=0.25}{v2,vc1}
\fmf{phantom,tension=0.5,right=0.25}{v2,vb2}
\fmf{plain,tension=0.5,left=0.25}{v3,vb2}
\fmf{plain,tension=0.5,left=0.25}{v4,vc3}
\fmf{plain,tension=0.5,right=0.25}{v5,vc3}
\fmf{phantom,tension=0.5,left=0.25}{v5,vb4}
\fmf{plain,tension=0.5,right=0.25}{v6,vb4}
\fmf{plain,tension=1.25,left=0}{vc1,vb1}
\fmf{plain,tension=1.25,left=0}{vb1,vb3}
\fmf{plain,tension=1.25,left=0}{vb3,vc3}
\fmf{plain,tension=1.25,left=0}{vb2,vb4}
\fmffreeze
\fmf{plain,tension=1,left=0}{vb1,vb2}
\fmf{plain,tension=1,left=0}{vb3,vb4}
\fmf{plain,tension=0.5,right=0,width=1mm}{v4,v6}
\fmffreeze
\fmfposition
}

\newcommand{\chionetwoonen}[1][black]{%
\fmftop{v1}
\fmfbottom{v4}
\fmfforce{(0.125w,h)}{v1}
\fmfforce{(0.125w,0)}{v4}
\fmffixed{(0.25w,0)}{v1,v2}
\fmffixed{(0.25w,0)}{v2,v3}
\fmffixed{(0.25w,0)}{v4,v5}
\fmffixed{(0.25w,0)}{v5,v6}
\fmffixed{(0,whatever)}{vc1,vc3}
\fmffixed{(0,whatever)}{vb2,vb4}
\fmffixed{(0,whatever)}{vc1,vb1}
\fmffixed{(0,whatever)}{vc1,vb3}
\fmffixed{(whatever,0.15h)}{vb2,vb1}
\fmffixed{(whatever,0.15h)}{vb3,vb4}
\fmf{plain,tension=0.5,right=0.25}{v1,vc1}
\fmf{plain,tension=0.5,left=0.25}{v2,vc1}
\fmf{phantom,tension=0.5,right=0.25}{v2,vb2}
\fmf{plain,tension=0.5,left=0.25}{v3,vb2}
\fmf{plain,tension=0.5,left=0.25}{v4,vc3}
\fmf{plain,tension=0.5,right=0.25}{v5,vc3}
\fmf{phantom,tension=0.5,left=0.25}{v5,vb4}
\fmf{plain,tension=0.5,right=0.25}{v6,vb4}
\fmf{plain,tension=1.25,left=0}{vc1,vb1}
\fmf{plain,tension=1.0,left=0}{vb2,vb4}
\fmf{plain,tension=1.25,left=0}{vb3,vc3}
\fmf{plain,tension=0.75,right=0.5}{vb1,vb3}
\fmffreeze
\fmf{plain,tension=1,left=0}{vb1,vb2}
\fmf{plain,tension=1,left=0}{vb3,vb4}
\fmf{plain,tension=0.5,right=0,width=1mm}{v4,v6}
\fmffreeze
\fmfposition
}

\newcommand{\chitwoonetwon}{%
\fmftop{v1}
\fmfbottom{v4}
\fmfforce{(0.125w,h)}{v1}
\fmfforce{(0.125w,0)}{v4}
\fmffixed{(0.25w,0)}{v1,v2}
\fmffixed{(0.25w,0)}{v2,v3}
\fmffixed{(0.25w,0)}{v4,v5}
\fmffixed{(0.25w,0)}{v5,v6}
\fmffixed{(0,whatever)}{vc1,vc3}
\fmffixed{(0,whatever)}{vb2,vb4}
\fmffixed{(0,whatever)}{vc1,vb1}
\fmffixed{(0,whatever)}{vc1,vb3}
\fmffixed{(whatever,0.15h)}{vb2,vb1}
\fmffixed{(whatever,0.15h)}{vb3,vb4}
\fmf{plain,tension=0.5,left=0.25}{v3,vc1}
\fmf{plain,tension=0.5,right=0.25}{v2,vc1}
\fmf{phantom,tension=0.5,left=0.25}{v2,vb2}
\fmf{plain,tension=0.5,right=0.25}{v1,vb2}
\fmf{plain,tension=0.5,right=0.25}{v6,vc3}
\fmf{plain,tension=0.5,left=0.25}{v5,vc3}
\fmf{phantom,tension=0.5,right=0.25}{v5,vb4}
\fmf{plain,tension=0.5,left=0.25}{v4,vb4}
\fmf{plain,tension=1.25,left=0}{vc1,vb1}
\fmf{plain,tension=0.75,left=0.5}{vb1,vb3}
\fmf{plain,tension=1.25,left=0}{vb3,vc3}
\fmf{plain,tension=1.0,right=0.0}{vb2,vb4}
\fmffreeze
\fmf{plain,tension=1,left=0}{vb1,vb2}
\fmf{plain,tension=1,left=0}{vb3,vb4}
\fmf{plain,tension=0.5,right=0,width=1mm}{v4,v6}
\fmffreeze
\fmfposition
}

\newcommand{\swfoneone}{%
\settoheight{\eqoff}{$\times$}%
\setlength{\eqoff}{0.5\eqoff}%
\addtolength{\eqoff}{-7.5\unitlength}%
\raisebox{\eqoff}{%
\fmfframe(1,0)(1,0){%
\begin{fmfchar*}(20,15)
\fmfleft{v1}
\fmfright{v2}
\fmffixed{(0.66w,0)}{vc1,vc2}
\fmf{plain}{v1,vc1}
\fmf{plain}{vc2,v2}
\fmf{plain,left=1}{vc1,vc2}
\fmf{plain,left=1}{vc2,vc1}
\fmffreeze
\fmfposition
\fmfipair{vm[]}
\svertex{vm1}{vpath(__vc1,__vc2)}
\svertex{vm2}{vpath(__vc2,__vc1)}
\end{fmfchar*}}}}

\newcommand{\swfoneonel}{%
\fmfleft{v1}
\fmfright{v2}
\fmffixed{(0.56w,0)}{vc1,vc2}
\fmf{plain}{v1,vc1}
\fmf{plain}{vc2,v2}
\fmf{plain,left=1}{vc1,vc2}
\fmf{plain,left=1}{vc2,vc1}
\fmffreeze
\fmfposition
\fmfipair{vm[]}
\svertex{vm1}{vpath(__vc1,__vc2)}
\svertex{vm2}{vpath(__vc2,__vc1)}
\svertex{vm3}{vpath(__v1,__vc1)}
\svertex{vm4}{vpath(__v2,__vc2)}
}

\newcommand{\swffourone}{%
\settoheight{\eqoff}{$\times$}%
\setlength{\eqoff}{0.5\eqoff}%
\addtolength{\eqoff}{-7.5\unitlength}%
\raisebox{\eqoff}{%
\fmfframe(1,0)(1,0){%
\begin{fmfchar*}(20,15)
\fmfleft{v1}
\fmfright{v2}
\fmffixed{(0.66w,0)}{vc1,vc2}
\fmffixed{(0.17w,0)}{vc1,vm1}
\fmffixed{(0.17w,0)}{vm2,vc2}
\fmf{plain}{v1,vc1}
\fmf{plain}{vc2,v2}
\fmf{plain,left=1}{vc1,vc2}
\fmf{plain,left=1}{vc2,vc1}
\fmffreeze
\fmf{plain}{vm1,vm2}
\fmfposition
\fmfipath{p[]}
\fmfipair{vu[],vv[]}
\fmfiset{p1}{vpath(__vc1,__vc2)}
\fmfiset{p2}{vpath(__vc2,__vc1)}
\dvertex{vu1}{vu2}{p1}
\dvertex{vv2}{vv1}{p2}
\fmfi{plain}{vu1--vv1}
\fmfi{plain}{vu2--vv2}
\fmfiset{p3}{vpath(__vu1,__vv1)}
\fmfiset{p4}{vpath(__vu2,__vv2)}
\end{fmfchar*}}}}

\newcommand{\swffouronel}{%
\fmfleft{v1}
\fmfright{v2}
\fmffixed{(0.56w,0)}{vc1,vc2}
\fmffixed{(0.14w,0)}{vc1,vm1}
\fmffixed{(0.14w,0)}{vm2,vc2}
\fmf{plain}{v1,vc1}
\fmf{plain}{vc2,v2}
\fmf{plain,left=1}{vc1,vc2}
\fmf{plain,left=1}{vc2,vc1}
\fmffreeze
\fmf{plain}{vm1,vm2}
\fmfposition
\fmfipath{p[]}
\fmfipair{vu[],vv[],vn[],vnn[],vt[]}
\fmfiset{p1}{vpath(__vc1,__vc2)}
\fmfiset{p2}{vpath(__vc2,__vc1)}
\fmfiset{p3}{vpath(__v1,__vc1)}
\fmfiset{p4}{vpath(__v2,__vc2)}
\dvertex{vu1}{vu2}{p1}
\dvertex{vv2}{vv1}{p2}
\fmfi{plain}{vu1--vv1}
\fmfi{plain}{vu2--vv2}
\svertex{vn1}{p3}
\svertex{vn2}{p4}
\dvertex{vnn1}{vnn2}{p3}
\dvertex{vnn3}{vnn4}{p4}
\svertex{vt2}{p1}
\fvertex{vt1}{1/6}{p1}
\fvertex{vt3}{5/6}{p1}
}

\newcommand{\gchionefourg}[1][black]{%
\fmftop{v1}
\fmfbottom{v5}
\fmfforce{(0.125w,h)}{v1}
\fmfforce{(0.125w,0)}{v5}
\fmffixed{(0.25w,0)}{v1,v2}
\fmffixed{(0.25w,0)}{v2,v3}
\fmffixed{(0.25w,0)}{v3,v4}
\fmffixed{(0.25w,0)}{v5,v6}
\fmffixed{(0.25w,0)}{v6,v7}
\fmffixed{(0.25w,0)}{v7,v8}
\fmf{plain,fore=#1}{v1,v5}
\fmf{plain,tension=0.5,right=0.25,fore=#1}{v2,vc1}
\fmf{plain,tension=0.5,left=0.25,fore=#1}{v3,vc1}
  \fmf{plain,tension=1.25,fore=#1}{vc1,vc2}
\fmf{plain,tension=0.5,left=0.25,fore=#1}{v6,vc2}
\fmf{plain,tension=0.5,right=0.25,fore=#1}{v7,vc2}
\fmf{plain,fore=#1}{v4,v8}
\fmf{plain,tension=0.5,right=0,width=1mm,fore=#1}{v5,v8}
\fmffixed{(0,.11h)}{vh1,vc2}
\fmffixed{(0,.22h)}{vh2,vc2}
\fmffreeze
\fmf{plain}{vh1,vh2}
\fmfposition
\fmfposition
\fmfipath{p[],pgl,pgr}
\fmfipair{vd[],vm[],vu[],vgd[],vglm[],vglu[],vgld[],vgl[],vgrm[],vgru[],vgrd[],vgr[],vfl[],vgfl[],vfr[],vgfr[]}
\fmfiset{p1}{vpath(__v2,__vc1)}
\fmfiset{p2}{vpath(__v3,__vc1)}
\fmfiset{p3}{vpath(__vc1,__vc2)}
\fmfiset{p5}{vpath(__v7,__vc2)}
\fmfiset{p4}{vpath(__v6,__vc2)}
\fmfiset{pgl}{vpath(__v1,__v5)}
\fmfiset{pgr}{vpath(__v4,__v8)}
\svertex{vm1}{p1}
\dvertex{vu1}{vd1}{p1}
\svertex{vm2}{p2}
\dvertex{vu2}{vd2}{p2}
\svertex{vm3}{p3}
\dvertex{vu3}{vd3}{p3}
\svertex{vm4}{p4}
\dvertex{vd4}{vu4}{p4}
\svertex{vm5}{p5}
\dvertex{vd5}{vu5}{p5}
\fvertex{vfr51}{1/6}{p5}
\fvertex{vfr55}{5/6}{p5}
\fvertex{vfl51}{1/6}{p4}
\fvertex{vfl55}{5/6}{p4}
\vvertex{vglu1}{vu1}{pgl}
\vvertex{vglm1}{vm1}{pgl}
\vvertex{vgld1}{vd1}{pgl}
\vvertex{vglu3}{vu3}{pgl}
\vvertex{vglm3}{vm3}{pgl}
\vvertex{vgld3}{vd3}{pgl}
\vvertex{vglu5}{vu5}{pgl}
\vvertex{vglm5}{vm5}{pgl}
\vvertex{vgld5}{vd5}{pgl}
\vvertex{vgl1}{vloc(__vc1)}{pgl}
\vvertex{vgl2}{vloc(__vc2)}{pgl}
\vvertex{vgru2}{vu2}{pgr}
\vvertex{vgrm2}{vm2}{pgr}
\vvertex{vgrd2}{vd2}{pgr}
\vvertex{vgru3}{vu3}{pgr}
\vvertex{vgrm3}{vm3}{pgr}
\vvertex{vgrd3}{vd3}{pgr}
\vvertex{vgfl51}{vfl51}{pgl}
\vvertex{vgfl55}{vfl55}{pgl}
\vvertex{vgfr51}{vfr51}{pgr}
\vvertex{vgfr55}{vfr55}{pgr}
\vvertex{vgru5}{vu5}{pgr}
\vvertex{vgrm5}{vm5}{pgr}
\vvertex{vgrd5}{vd5}{pgr}
\vvertex{vgr1}{vloc(__vc1)}{pgr}
\vvertex{vgr2}{vloc(__vc2)}{pgr}
\fmfi{plain}{vd4--vd5}
\fmfi{plain}{vu4--vu5}
}

\newcommand{\threetotwo}[1][black]{%
\fmftop{v1}
\fmfbottom{v3}
\fmfforce{(0.2w,h)}{v1}
\fmfforce{(0.125w,0)}{v3}
\fmffixed{(0.6w,0)}{v1,v2}
\fmffixed{(0.375w,0)}{v3,v4}
\fmffixed{(0.375w,0)}{v4,v5}
\fmffixed{(0,0.5h)}{v4,vc1}
\fmf{plain,tension=0.5,left=0.3,fore=#1}{v3,vc1}
\fmf{plain,tension=0.5,right=0.3,fore=#1}{v5,vc1}
\fmf{plain,tension=0.5,right=0,width=1mm,fore=#1}{v3,v5}
\fmf{plain,fore=#1}{v4,vc1}
\fmffreeze
\fmfposition
\fmfipath{p[]}
\fmfipair{vd[],vc[],vs[]}
\fmfiset{p1}{vpath(__v3,__vc1)}
\fmfiset{p2}{vpath(__v5,__vc1)}
\fmfiset{vs1}{vloc(__v1)}
\fmfiset{vs2}{vloc(__v2)}
\fvertex{vd1}{3/4}{p1}
\fvertex{vd2}{3/4}{p2}
\fvertex{vc1}{1/2}{p2}
\fmfi{photon}{vs1--vd1}
\fmfi{photon}{vs2--vd2}
}

\newcommand{\ctprop}[1][black]{%
\fmfleft{v1}
\fmfright{v2}
\fmf{plain,tension=0.5,fore=#1}{v1,vc1}
\fmf{plain,tension=0.5,fore=#1}{v2,vc1}
\fmf{phantom,tension=0,fore=#1}{v1,v2}
\fmfv{decor.shape=diamond,decor.filled=full, decor.size=4thick}{vc1}
\fmfposition
\fmfipath{p[]}
\fmfipair{vd[],vm[],vu[]}
\fmfiset{p1}{vpath(__v1,__vc1)}
\fmfiset{p2}{vpath(__v2,__vc1)}
\fmfiset{p3}{vpath(__v1,__v2)}
\svertex{vm1}{p1}
\svertex{vm2}{p2}
\svertex{vm3}{p3}
}

\newcommand{\ctone}[1][black]{%
\fmftop{v1}
\fmfbottom{v3}
\fmfforce{(0.125w,h)}{v1}
\fmfforce{(0.125w,0)}{v3}
\fmffixed{(0.3w,0)}{v1,v2}
\fmffixed{(0.15w,0)}{v3,v4}
\fmffixed{(0.15w,0)}{v4,v5}
\fmf{plain,tension=0.5,right=0.25,fore=#1}{v1,vc1}
\fmf{plain,tension=0.5,left=0.25,fore=#1}{v2,vc1}
\fmf{plain,tension=1.25,fore=#1}{vc1,vc2}
\fmf{plain,tension=1.0,left=0.,fore=#1}{v4,vc2}
\fmf{plain,tension=0.5,right=0,width=1mm,fore=#1}{v3,v5}
\fmfv{decor.shape=diamond,decor.filled=full, decor.size=7thick}{v4}
\fmfposition
\fmfipath{p[]}
\fmfipair{vd[],vm[],vu[]}
\fmfiset{p1}{vpath(__v1,__vc1)}
\fmfiset{p2}{vpath(__v2,__vc1)}
\fmfiset{p3}{vpath(__vc1,__vc2)}
\fmfiset{p4}{vpath(__v4,__vc2)}
\fmfiset{p5}{vpath(__v3,__v5)}
\svertex{vm1}{p1}
\dvertex{vu1}{vd1}{p1}
\svertex{vm2}{p2}
\dvertex{vu2}{vd2}{p2}
\svertex{vm3}{p3}
\dvertex{vu3}{vd3}{p3}
\svertex{vm4}{p4}
\dvertex{vd4}{vu4}{p4}
\svertex{vm5}{p5}
}

\begin{document}

\begin{fmffile}{graphs}
\fmfcmd{%
input Dalgebra
}

\fmfcmd{%
def getmid(suffix p) =
  pair p.mid[], p.off[], p.dir[];
  for i=0 upto 36:
    p.dir[i] = dir(5*i);
    p.mid[i]+p.off[i] = directionpoint p.dir[i] of p;
    p.mid[i]-p.off[i] = directionpoint -p.dir[i] of p;
  endfor
enddef;
}

\fmfcmd{%
marksize=2mm;
def draw_mark(expr p,a) =
  begingroup
    save t,tip,dma,dmb; pair tip,dma,dmb;
    t=arctime a of p;
    tip =marksize*unitvector direction t of p;
    dma =marksize*unitvector direction t of p rotated -45;
    dmb =marksize*unitvector direction t of p rotated 45;
    linejoin:=beveled;
    draw (-.5dma.. .5tip-- -.5dmb) shifted point t of p;
  endgroup
enddef;
style_def derplain expr p =
    save amid;
    amid=.5*arclength p;
    draw_mark(p, amid);
    draw p;
enddef;
style_def derphoton expr p =
    save amid;
    amid=.5*arclength p;
    draw_mark(p, amid);
    draw wiggly p;
enddef;
def draw_marks(expr p,a) =
  begingroup
    save t,tip,dma,dmb,dmo; pair tip,dma,dmb,dmo;
    t=arctime a of p;
    tip =marksize*unitvector direction t of p;
    dma =marksize*unitvector direction t of p rotated -45;
    dmb =marksize*unitvector direction t of p rotated 45;
    dmo =marksize*unitvector direction t of p rotated 90;
    linejoin:=beveled;
    draw (-.5dma.. .5tip-- -.5dmb) shifted point t of p withcolor 0white;
    draw (-.5dmo.. .5dmo) shifted point t of p;
  endgroup
enddef;
style_def derplains expr p =
    save amid;
    amid=.5*arclength p;
    draw_marks(p, amid);
    draw p;
enddef;
def draw_markss(expr p,a) =
  begingroup
    save t,tip,dma,dmb,dmo; pair tip,dma,dmb,dmo;
    t=arctime a of p;
    tip =marksize*unitvector direction t of p;
    dma =marksize*unitvector direction t of p rotated -45;
    dmb =marksize*unitvector direction t of p rotated 45;
    dmo =marksize*unitvector direction t of p rotated 90;
    linejoin:=beveled;
    draw (-.5dma.. .5tip-- -.5dmb) shifted point t of p withcolor 0white;
    draw (-.5dmo.. .5dmo) shifted point arctime a+0.25 mm of p of p;
    draw (-.5dmo.. .5dmo) shifted point arctime a-0.25 mm of p of p;
  endgroup
enddef;
style_def derplainss expr p =
    save amid;
    amid=.5*arclength p;
    draw_markss(p, amid);
    draw p;
enddef;
style_def dblderplains expr p =
    save amidm;
    save amidp;
    amidm=.5*arclength p-0.75mm;
    amidp=.5*arclength p+0.75mm;
    draw_mark(p, amidm);
    draw_marks(p, amidp);
    draw p;
enddef;
style_def dblderplainss expr p =
    save amidm;
    save amidp;
    amidm=.5*arclength p-0.75mm;
    amidp=.5*arclength p+0.75mm;
    draw_mark(p, amidm);
    draw_markss(p, amidp);
    draw p;
enddef;
style_def dblderplainsss expr p =
    save amidm;
    save amidp;
    amidm=.5*arclength p-0.75mm;
    amidp=.5*arclength p+0.75mm;
    draw_marks(p, amidm);
    draw_markss(p, amidp);
    draw p;
enddef;
}


\thispagestyle{empty}
\begin{flushright}\footnotesize
\texttt{UUITP-23/11} 
\vspace{0.8cm}\end{flushright}

\begin{center}
{\Large\textbf{\mathversion{bold} Supergraphs and the cubic Leigh-Strassler model}}

\vspace{1.5cm}

\textrm{Joseph~A.~Minahan}
\vspace{8mm}

\textit{ Department of Physics and Astronomy\\ Uppsala University\\ Box 520\\
SE-751 20 Uppsala, Sweden}\\
\texttt{joseph.minahan@fysast.uu.se} \vspace{3mm}

\vspace{3mm}


\par\vspace{1cm}

\textbf{Abstract} \vspace{5mm}

\begin{minipage}{14cm}

We discuss supergraphs and their relation to  ``chiral functions" in $\NN=4$ Super Yang-Mills.  Based on the magnon dispersion relation and an explicit three-loop result of Sieg's we make an all loop conjecture for the rational contributions of certain classes of supergraphs.  We then apply superspace techniques  to the ``cubic" branch of Leigh-Strassler $\NN=1$ superconformal theories.  We show that there are order $2^L/L$ single trace operators of length $L$ which have zero anomalous dimensions to all loop order in the planar limit.  We then  compute the anomalous dimensions for another class of single trace operators we call one-pair states.  Using the conjecture we can find a simple expression for the 
rational part of the 
anomalous dimension which we argue is valid  at least  up to and including five-loop order.  Based on an explicit computation we can compute the anomalous dimension for these operators to four loops.
\end{minipage}

\end{center}

\vspace{0.5cm}


\newpage
\setcounter{page}{1}
\renewcommand{\thefootnote}{\arabic{footnote}}
\setcounter{footnote}{0}
\setcounter{equation}{0}

\tableofcontents
\section{Introduction}

Integrability has proven to be a powerful tool in computing the spectrum of operators in highly superconformal gauge theories (see \cite{Beisert:2010jr} for a comprehensive review.)  The most studied theory is of course $\NN=4$ Super Yang-Mills (SYM).  A while ago, Leigh and Strassler showed that $\NN=4$ SYM has three independent deformations that preserve $\NN=1$ superconformal symmetry 
 \cite{Leigh:1995ep}.  The Leigh-Strassler superpotential can be written as
\be\label{superpot}
W=\kappa\left(\Tr(XYZ-q XZY)+\frac{h}{3}(\Tr X^3+\Tr Y^3+\Tr Z^3)\right)\,.
\ee
By a chiral field phase rotation  $\kappa$  can be chosen real.   There is a further relation of the couplings to $g_\YM$, the Yang-Mills coupling, leaving a  three complex dimensional space of $\NN=1$ superconformal theories.

Leigh and Strassler's argument is essentially an existence proof for the  marginal directions, but  the relation of the couplings is known, at least to the first few loop orders,  and in the planar limit is given by \cite{Aharony:2002tp}
\be\label{couprel}
2g_\YM^2=\kappa\bar\kappa(1+q\bar q+h\bar h)\,.
\ee
  The most well known deformation is the so-called $\beta$-deformation where $h=0$ and $q=e^{-2i\pi\beta}$ with $\beta$  real (see \cite{Zoubos:2010kh} for a review).  After a field redefinition, this modifies the $\NN=4$ superpotential  to 
\be
W=g_\YM\Tr(X[Y,Z])\to g_\YM\Tr(e^{ i\pi \beta}XYZ-e^{- i\pi\beta}XZY)\,.
\ee
Staying in the planar limit,  the relation in (\ref{couprel}) is exact \cite{Khoze:2005nd,Mauri:2005pa}.  Furthermore, the computation for the spectrum of local operators is an integrable problem \cite{Roiban:2003dw,Berenstein:2004ys,Beisert:2005if,Frolov:2005ty}.  In \cite{Berenstein:2004ys} it was shown at the one-loop level that the corresponding Bethe equations are the same as $\NN=4$ except for a $\beta$-dependent shift.  This was extended to all loops in \cite{Beisert:2005if}.  The supergravity dual for the $\beta$-deformed theory is known \cite{Lunin:2005jy} and its world-sheet theory has been shown to be classically integrable, even though it is not a coset \cite{Frolov:2005dj,Frolov:2005ty}.

For a general deformation it appears that integrability might be lost. Even at the one-loop level, for complex $\beta$ the resulting spin-chain is a known integrable model only in the $SU(2)$ sector\cite{Berenstein:2004ys}.  Moreover, the relation between the couplings in (\ref{couprel}) gets higher loop corrections starting at four-loop order \cite{Khoze:2005nd,Elmetti:2006gr,Elmetti:2007up}.   

Less attention has been paid to  deformations with nonzero $h$, although  interesting results have still been found.  
  In
\cite{Bundzik:2005zg,Mansson:2007sh,Mansson:2008xv} it has been   shown that for the choice $q=(1+\rho)e^{\pi i n/3}$, $h=\rho e^{\pi im/3}$ where $\rho$ is real and $n$ and $m$ are integers, the $R$-matrix in the $SU(3)$ sector satisfies the Yang-Baxter equation.  However, it was also shown that all such theories are related to the $\beta$-deformed theories via similarity transformations \cite{Bundzik:2005zg}.   Another interesting branch has $q=0$ and $|h|=1$, which was shown to be integrable at the one-loop level \cite{Mansson:2008xv} and also has been conjectured to require no higher-loop corrections to (\ref{couprel}) \cite{Bork:2007bj}.

In this paper we consider a branch of Leigh-Strassler theories that is far away from the $\NN=4$ branch.  Here we let $\kappa=0$ in (\ref{superpot}) while at the same time taking $h\to\infty$ such that $\hh\equiv h\kappa$ is finite.  The superpotential then takes the form
\be\label{Z3superpot}
W=\frac{\hh}{3}\big(\Tr X^3+\Tr Y^3+\Tr Z^3\big)\,,
\ee
where $\hh$ can be chosen real.
We call this the ``cubic" Leigh-Strassler  model\footnote{Of course, the entire Leigh-Strassler moduli space is cubic.  The name chosen here is intended to be a shortened version of Fermat cubic.}.  The model has a  $U(1)$ $R$-symmetry, but 
unlike the $\beta$-deformed theory, it does not have an extra $U(1)\times U(1)$ global symmetry.  It does have an $S_3\times Z_3\times Z_3$ discrete symmetry, which is a larger symmetry group than the generic superpotential in (\ref{superpot}), which only has a $Z_3\times Z_3$ symmetry.  

Using the relations of the couplings in (\ref{couprel}), we see  that for the cubic model
$|\hh|^2=2g_\YM^2$ in the planar limit, at least to the lowest few loop orders.  Starting at four-loops this relation gets corrected \cite{Bork:2007bj}.  Given the nature of the superpotential in (\ref{Z3superpot}) many single trace operators composed of scalar fields do not mix with other operators.  In fact, we will show that an exponentially large number, $\sim 2^L/L$ where $L$ is the length of the operator,  receive no anomalous dimensions to any loop order in the planar limit.  We will refer to such states as  planar protected.  This suggests that the gravity dual for this theory is stringy in nature.

This result about planar protected states relies on  properties of supergraphs \cite{Gates:1983nr} (see \cite{Sieg:2010jt} for a review that centers on supergraphs in $\NN=4$ SYM.)  Supergraph technology can vastly reduce the number of Feynman diagrams that  appear in an anomalous dimension calculation.  In the case of the $SU(2)$ sector of $\NN=4$ SYM, the one-loop diagrams in \cite{Minahan:2002ve} are reduced to a single diagram using supergraphs.  At higher loops, the heroic effort to compute the four-loop wrapping corrections in \cite{Fiamberti:2007rj,Fiamberti:2008sh} would have been well-nigh impossible without using supergraphs.  Even in ABJM/ABJ theories, the four-loop calculations in \cite{Minahan:2009aq,Minahan:2009wg} to compute corrections to an unknown function of the 't Hooft parameters were  simplified in \cite{Leoni:2010tb} by using $\NN=2$ superspace \cite{Benna:2008zy}.  In both $\NN=4$ SYM and ABJM/ABJ, a key to the simplification is a set of finiteness conditions that can be used to discard many graphs \cite{Sieg:2010tz}.

An important property of the supergraphs is that they naturally come with factors of  so-called chiral functions (to be explained in the text).   The chiral functions can  be read off from the ``skeletal" part of the supergraph, which is the supergraph with all vector propagators and vertices removed.  In the $SU(2)$ sector many of the chiral functions are equivalent  and one ends up combining the different classes of supergraphs together to, say, compare a supergraph calculation with a result expected from the magnon dispersion relation \cite{Sieg:2010tz}.

In this paper we offer a conjecture that disentangles  the 
rational
 contributions of certain chiral functions to the magnon dispersion relation.  It is based on  Sieg's explicit three-loop computation  \cite{Sieg:2010tz} of the dilation operator in the $SU(2)$ sector \cite{Beisert:2004hm}. It proposes a natural decomposition of  individual contributions which are indistinguishable in this 
sector, but act differently outside of it.  This allows us to find an all-loop contribution
 for the rational part 
of supergraphs with a particular chiral function.    
This technique does not work for all chiral functions, but only for a class of them that we call connected.

The chiral functions arise out of the $\NN=4$ superpotential, but given the all-loop contribution of supergraphs for the connected chiral functions, we can apply it to the $\NN=1$ superconformal deformations by replacing the $\NN=4$ chiral functions with those appropriate for the $\NN=1$ theory.   In the $\beta$-deformed case this is straightforward \cite{Fiamberti:2008sm,Fiamberti:2008sn}, but for the cubic model this is  more complicated because of the four-loop correction to $\hh$.  The correction implies that we must include  additional supergraphs in the overall computation, and depending on one's prescription, infinite counterterms. 
Nevertheless, we will argue that for a particular class of single trace operators the tuning of $\hh$ that keeps the cubic model superconformal also cancels out the contributions of the  additional supergraphs,  at least to five loops.    For these states only one chiral function acts nontrivially and we use its form for the cubic model to explicitly find the anomalous dimension of these operators.   Starting at six loops there are other types of these extra diagrams that we have not yet determined if they will cancel with the tuning.

While we claim that the conjecture is valid for the rational contributions, starting at four loops there are higher transcendental contributions that do not cancel out for the individual chiral functions, but do cancel once identifications are made between chiral functions \cite{MinSieg}.  Nonetheless, certain chiral functions are easier to compute than others, and one can use the $\NN=4$ dispersion relation to extract the contribution of the more difficult to compute chiral function by directly computing the easier one.  We have done this at the four-loop level  \cite{MinSieg} to find the chiral function relevant for our $\NN=1$ theory and can thereby find the anomalous dimensions to four-loop order for this class of operators. 

In section 2 we review some relevant properties of superspace diagrams.  In section 3 we make the conjecture relating the rational part of superspace diagrams to the dilatation operator.  In section 4 we apply the 
results of sections 2 and 3 to the cubic model.  We first show how to take the corrections of $\hh$ into account.  We then present the one-loop Hamiltonian in the chiral scalar sector, writing it as an $SU(3)$ analog of the anti-ferromagnetic Ising model.    At higher loops, aside from arguing the claims of the preceding paragraphs, we also briefly discuss some other operators outside of these classes.  We also discuss the possibility of a finite Hagedorn temperature at strong coupling.  In section 5 we summarize our results and discuss future directions. 

\section{Superspace for composite operators}

In this section we discuss the correlators of composite  operators in superspace \cite{Gates:1983nr,Sieg:2010jt}.  For the purposes of this discussion, a composite operator $\OO$ in $\NN=4$ SYM and their $\NN=1$ deformations will mean a single trace operator containing the scalar chiral fields $X$, $Y$ and $Z$, which we also write as $\phi^i$, $i=1,2,3$, as well as the gaugino chiral field $W_\al=-\frac{1}{4}\bar D^2D_\al V$, where $V$ is the vector superfield.  The anti-chiral fields are given by $\bar\phi_i$ and $\bar W_{\dot \al}$.  In $\NN=4$ language, these are the operators that make up the closed $SU(2|3)$ sector \cite{Beisert:2003ys}.  We will also consider scalar composites where only the scalar fields are present.

The building blocks of a supergraph are the propagators for the chiral and the vector fields, and the vertices.  A chiral field progagator is denoted by 
\begin{eqnarray}
\langle\,{{\phi^i\,}^a}_b\,{\bar{\phi_j}^c}_{\,d}\,\rangle
&&=
\settoheight{\eqoff}{$\times$}%
\setlength{\eqoff}{0.5\eqoff}%
\addtolength{\eqoff}{-3.75\unitlength}%
\raisebox{\eqoff}{%
\fmfframe(1,0)(1,0){%
\begin{fmfchar*}(15,7.5)
\fmfleft{v1}
\fmfright{v2}
\fmfforce{0.0625w,0.5h}{v1}
\fmfforce{0.9375w,0.5h}{v2}
\fmf{plain}{v1,v2}
\fmffreeze
\fmfposition
\fmfipath{pm[]}
\fmfiset{pm1}{vpath(__v1,__v2)}
\nvml{1}{$\scriptstyle p$}
\end{fmfchar*}}}
=
{\delta^i}_j\frac{{\delta^a}_d{\delta^c}_b}{p^2}\delta^4(\theta_1-\theta_2)
\end{eqnarray}
where the raised  indices $a$, $c$ refer to fundamental gauge indices and the lowered  indices $b$, $d$ refer to anti-fundamental gauge indices.  Strictly speaking this is the propagator for a $U(N)$ gauge theory, but this is immaterial in the large $N$ limit.  Likewise, the propagator for the vector fields is
\begin{eqnarray}
\langle\, {V^a}_b\,{V^c}_d\,\rangle
&&=
\settoheight{\eqoff}{$\times$}%
\setlength{\eqoff}{0.5\eqoff}%
\addtolength{\eqoff}{-3.75\unitlength}%
\raisebox{\eqoff}{%
\fmfframe(1,0)(1,0){%
\begin{fmfchar*}(15,7.5)
\fmfleft{v1}
\fmfright{v2}
\fmfforce{0.0625w,0.5h}{v1}
\fmfforce{0.9375w,0.5h}{v2}
\fmf{photon}{v1,v2}
\fmffreeze
\fmfposition
\fmfipath{pm[]}
\fmfiset{pm1}{vpath(__v1,__v2)}
\nvml{1}{$\scriptstyle p$}
\end{fmfchar*}}}
=-\frac{{\delta^a}_d{\delta^c}_b}{p^2}\delta^4(\theta_1-\theta_2)
\end{eqnarray}
Note that the superspace propagators have dimension $-4$.  

The possible vertices include the chiral vertex,
\begin{eqnarray}
V_{\phi^i\phi^j\phi^k}
&=\cvert{plain}{plain}{plain}{}{$\scriptstyle\barD^2$}{$\scriptstyle\barD^2$}
i\,\left(G_{ijk}{\delta^{a_2}}_{b_1}{\delta^{a_3}}_{b_2}{\delta^{a_1}}_{b_3}+G_{ikj}{\delta^{a_3}}_{b_1}{\delta^{a_1}}_{b_2}{\delta^{a_2}}_{b_3}\right)\,
\end{eqnarray}
where $G_{ijk}$ is a coupling that depends on the relevant theory.  In the case of $\NN=4$ SYM we have that $G_{ijk}=g_{\rm YM}\,\eps_{ijk}$.  The gauge indices are numbered in the order that the fields appear in the vertex. Two of the three legs of the vertex include the square of super-derivatives $\barD^2=\eps^{\dot\alpha\dot\beta}\barD_{\dot\al}\barD_{\dot\beta}$.  Since the superpotential has no mass terms,  a chiral vertex cannot connect with another chiral vertex.  

There is also the anti-chiral vertex
\begin{eqnarray}
V_{\bar\phi_i\bar\phi_j\bar\phi_k}
&&=\cvert{plain}{plain}{plain}{}{$\scriptstyle\D^2$}{$\scriptstyle\D^2$}
i\,\left(G^{ijk}{\delta^{a_2}}_{b_1}{\delta^{a_3}}_{b_2}{\delta^{a_1}}_{b_3}+G^{ikj}{\delta^{a_3}}_{b_1}{\delta^{a_1}}_{b_2}{\delta^{a_2}}_{b_3}\right)\,,
\end{eqnarray}
which can connect with a chiral vertex.  
Other vertices which can appear involve the gauge fields.  The ones relevant for  three-loop calculations are
\begin{eqnarray}
V_{\bar\phi_j V\phi^i}
&=&
\cvert{photon}{plain}{plain}{}{$\scriptstyle\barD^2$}{$\scriptstyle\D^2$}
g_\YM\delta_i^j\left({\delta^{a_2}}_{b_1}{\delta^{a_3}}_{b_2}{\delta^{a_1}}_{b_3}-{\delta^{a_3}}_{b_1}{\delta^{a_1}}_{b_2}{\delta^{a_2}}_{b_3}\right)\nn\\
V_{V^2\bar\phi_i\phi^j}
&=&
\qvert{photon}{photon}{plain}{plain}{}{}{$\scriptstyle\D^2$}{$\scriptstyle\barD^2$}\frac{g_\YM^2}{2}\delta_j^i\left({\delta^{a_2}}_{b_1}{\delta^{a_3}}_{b_2}{\delta^{a_4}}_{b_3}{\delta^{a_1}}_{b_4}+{\delta^{a_1}}_{b_2}{\delta^{a_3}}_{b_1}{\delta^{a_4}}_{b_3}{\delta^{a_2}}_{b_4}\right)\nn
\\
V_{V^2\phi^i\bar\phi_j}
&=&
\qvert{photon}{photon}{plain}{plain}{}{}{$\scriptstyle\barD^2$}{$\scriptstyle\D^2$}\frac{g_\YM^2}{2}\delta_i^j\left({\delta^{a_2}}_{b_1}{\delta^{a_3}}_{b_2}{\delta^{a_4}}_{b_3}{\delta^{a_1}}_{b_4}+{\delta^{a_1}}_{b_2}{\delta^{a_3}}_{b_1}{\delta^{a_4}}_{b_3}{\delta^{a_2}}_{b_4}\right)\nn
\\
V_{V\bar\phi_iV\phi^j}
&=&
\qvert{photon}{plain}{photon}{plain}{}{$\scriptstyle\D^2$}{}{$\scriptstyle\barD^2$}(-g_\YM^2)\delta_j^i\left({\delta^{a_2}}_{b_1}{\delta^{a_3}}_{b_2}{\delta^{a_4}}_{b_3}{\delta^{a_1}}_{b_4}\right)
\end{eqnarray}
Note that all of the above vertices come with an integral over superspace $\int d^4\theta$.  

Finally, the composite operator itself can be thought of as a chiral vertex.  If it is a scalar composite with $L$ scalar chiral fields, then we denote it as a chiral vertex  with $L$ legs,
\begin{eqnarray}
{\OO=\dots}
\settoheight{\eqoff}{$\times$}%
\setlength{\eqoff}{0.5\eqoff}%
\addtolength{\eqoff}{-8.5\unitlength}%
{%
\raisebox{\eqoff}{%
\fmfframe(-1,1)(4,1){%
\begin{fmfchar*}(20,15)
\fmftop{v5}
\fmfbottom{v6}
\fmfforce{(0.125w,h)}{v5}
\fmfforce{(0.125w,0)}{v6}
\fmffixed{(0.25w,0)}{v2,v1}
\fmffixed{(0.25w,0)}{v3,v2}
\fmffixed{(0.25w,0)}{v4,v3}
\fmffixed{(0.25w,0)}{v5,v4}
\fmffixed{(0.25w,0)}{v6,v7}
\fmffixed{(0.25w,0)}{v7,v8}
\fmffixed{(0.25w,0)}{v8,v9}
\fmffixed{(0.25w,0)}{v9,v10}
\fmf{plain}{v1,v10}
\fmf{plain}{v2,v9}
\fmf{plain}{v3,v8}
\fmf{plain}{v4,v7}
\fmf{plain}{v5,v6}
\fmffreeze
\fmf{plain,tension=1,left=0,width=1mm}{v6,v10}
\fmfposition
\fmfipath{p[]}
\fmfiset{p1}{vpath(__v1,__v10)}
\fmfiset{p2}{vpath(__v2,__v9)}
\fmfiset{p3}{vpath(__v3,__v8)}
\fmfiset{p4}{vpath(__v4,__v7)}
\fmfiset{p5}{vpath(__v5,__v6)}
\fmfis{phantom,ptext.clen=6,ptext.hout=3,ptext.oout=12,ptext.out=$\scriptstyle\barD^2$,ptext.sep=;}{p2}
\fmfis{phantom,ptext.clen=6,ptext.hout=3,ptext.oout=12,ptext.out=$\scriptstyle\barD^2$,ptext.sep=;}{p1}
\fmfis{phantom,ptext.clen=6,ptext.hout=3,ptext.oout=12,ptext.out=$\scriptstyle\barD^2$,ptext.sep=;}{p4}
\fmfis{phantom,ptext.clen=6,ptext.hout=3,ptext.oout=12,ptext.out=$\scriptstyle\barD^2$,ptext.sep=;}{p5}
\end{fmfchar*}}}}
\ \ \dots
\end{eqnarray}
Like the three-point chiral vertex, all but one leg coming out of the composite operator has a $\barD^2$ attached to the leg.  Integrating by parts, it is possible to choose any of the legs emerging from the operator to be the one without the $\barD^2$ factor.  If we were to consider the more general case of composites that also contain $W_\al$ fields, then the composite vertex would like 
\begin{eqnarray}
{\OO=\dots}
\settoheight{\eqoff}{$\times$}%
\setlength{\eqoff}{0.5\eqoff}%
\addtolength{\eqoff}{-8.5\unitlength}%
{%
\raisebox{\eqoff}{%
\fmfframe(-1,1)(4,1){%
\begin{fmfchar*}(20,15)
\fmftop{v5}
\fmfbottom{v6}
\fmfforce{(0.125w,h)}{v5}
\fmfforce{(0.125w,0)}{v6}
\fmffixed{(0.25w,0)}{v2,v1}
\fmffixed{(0.25w,0)}{v3,v2}
\fmffixed{(0.25w,0)}{v4,v3}
\fmffixed{(0.25w,0)}{v5,v4}
\fmffixed{(0.25w,0)}{v6,v7}
\fmffixed{(0.25w,0)}{v7,v8}
\fmffixed{(0.25w,0)}{v8,v9}
\fmffixed{(0.25w,0)}{v9,v10}
\fmf{plain}{v1,v10}
\fmf{plain}{v2,v9}
\fmf{photon}{v3,v8}
\fmf{plain}{v4,v7}
\fmf{photon}{v5,v6}
\fmffreeze
\fmf{plain,tension=1,left=0,width=1mm}{v6,v10}
\fmfposition
\fmfipath{p[]}
\fmfiset{p1}{vpath(__v1,__v10)}
\fmfiset{p2}{vpath(__v2,__v9)}
\fmfiset{p3}{vpath(__v3,__v8)}
\fmfiset{p4}{vpath(__v4,__v7)}
\fmfiset{p5}{vpath(__v5,__v6)}
\fmfis{phantom,ptext.clen=6,ptext.hout=3,ptext.oout=12,ptext.out=$\scriptstyle\barD^2$,ptext.sep=;}{p2}
\fmfis{phantom,ptext.clen=6,ptext.hout=3,ptext.oout=12,ptext.out=$\scriptstyle\barD^2$,ptext.sep=;}{p1}
\fmfis{phantom,ptext.clen=5,ptext.hout=5,ptext.oout=15,ptext.out=$\scriptstyle-\frac{1}{4}\D_\beta$,ptext.sep=;}{p3}
\fmfis{phantom,ptext.clen=6,ptext.hout=3,ptext.oout=12,ptext.out=$\scriptstyle\barD^2$,ptext.sep=;}{p4}
\fmfis{phantom,ptext.clen=5,ptext.hout=5,ptext.oout=20,ptext.out=$\scriptstyle-\frac{1}{4}\barD^2D_\al$,ptext.sep=;}{p5}
\end{fmfchar*}}}}
\ \ \dots
\end{eqnarray}
Again, one leg is missing a $\bar D^2$, in this case it is a  $W_\al$ leg.

As is well known, the dimension of $\OO$ has loop corrections, and these loops lead to mixing among the operators. It is hence best to consider the dilatation operator $\DD$ that acts on the composite operator and whose eigenvalues are the dimensions of linear combinations of composite operators.  $\DD$ has a perturbative expansion in $g_{\rm YM}^2$, but  in the large $N$ limit, one only need consider planar contributions to $\DD$.  Hence, the dilation operator can be written as 
\begin{eqnarray}
\DD=\sum_{n=0}^\infty g^{2n}\DD_n,
\end{eqnarray}
where $g^2\equiv\frac{g_{\rm YM}^2N}{16\pi^2}=\frac{\la}{16\pi^2}$.    Focusing on the case of scalar composites, the lowest order contribution  is just $\DD_0=L$.  The first order contribution comes from the supergraph 
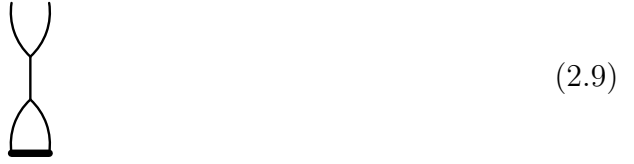
\begin{equation}
\begin{aligned}\label{Z1}
\settoheight{\eqoff}{$\times$}%
\setlength{\eqoff}{0.5\eqoff}%
\addtolength{\eqoff}{-11\unitlength}%
\raisebox{\eqoff}{%
\fmfframe(-1,1)(-11,1){%
\begin{fmfchar*}(20,20)
\chione
\end{fmfchar*}}}
\end{aligned}
\end{equation}
which is a one-loop supergraph with a chiral and an anti-chiral vertex.

The evaluation of any supergraph requires the $D$-algebra \cite{Gates:1983nr,Fiamberti:2008sh,Sieg:2010tz}, which combines integration by parts with identities for the super-derivatives.   Applying this to the supergraph in (\ref{Z1}), we can choose to have one of the propagators in the loop with a $D^2$ and $\barD^2$ factor and the other propagator with no super-derivatives.  The $D^2$ and $\barD^2$ cancel one of the $\delta^4(\theta)$ terms from the propagators while the other $\delta^4(\theta)$ term is absorbed by the integral over superspace from the vertex, leaving an ordinary scalar integral.  The supergraph will also come with a factor of $G_{ijk}G^{klm}$, where $ij$ are the two incoming indices of the scalar fields (going left to right) and $lm$ are the outgoing indices (going right to left).  

Before stating the result of this integration, let us do a simple power counting exercise for the supergraph.  
The loop has one vertex (hence one integration over $\int d^4\theta$), two propagators , 4 super-derivatives and one loop momentum integral.  Hence, the dimension is $2\!-\!8\!+\!2\!+\!4=0$.  Thus we expect this integral to be logarithmically divergent and lead to a contribution to the anomalous dimension.  In fact it does.  The divergent part of the scalar integral is
$\frac{1}{32\pi^2\eps}G_{ijk}G^{klm}$ where we have dimensionally reduced to $D=4-2\eps$ \cite{Sieg:2010tz}.  If we consider the contribution in the $SU(2)$ sector for $\NN=4$ SYM, where the scalar fields are restricted to $X$ and $Y$, then the dilatation operator becomes
\be\label{D1}
\DD_1=2\,\chi(1)\,,
\ee
where $\chi(1)$ is an example of a  chiral function.
The chiral functions are defined as \cite{Fiamberti:2008sh,Sieg:2010tz}\footnote{Note that our convention differs by a factor of $(-1)^n$.}
\be
\chi(a_1,a_2\dots a_n)\equiv\sum_{j=0}^{L-1}(1-P_{j+a_1,j+a_1+1})(1-P_{j+a_2,j+a_2+1})\dots(1-P_{j+a_n,j+a_n+1})\,,
\ee
where $P_{j,j+1}$ is the exchange operator between neighboring sites.  A useful property of chiral functions is 
\be\label{prop1}
\chi(a_1,\dots a_j,a_{j+1}\dots a_n)=\chi(a_1,\dots a_{j+1},a_j\dots a_n)
\ee
if $|a_j-a_{j+1}|\ge 2$, while
\be\label{prop2}
\chi(a_1,\dots a_j,a_{j+1}\dots a_n)=2\,\chi(a_1,\dots a_j\dots a_n)
\ee
if $a_j=a_{j+1}$.  We will call a chiral function connected if it can be arranged using  (\ref{prop2}) to a chiral function where $|a_j-a_{j+1}|=1$ for all $j$.  Otherwise, the chiral function is disconnected.

Power counting can be used to exclude many supergraphs \cite{Fiamberti:2008sh,Sieg:2010tz}.  One of the most powerful rules is that in order to have a divergent supergraph, there must be an anti-chiral vertex that sits outside of all loops\cite{Sieg:2010tz}.  Clearly the supergraph in (\ref{Z1}) satisfies this criterion.  But then it must also follow that if a supergraph has  its composite operator legs connected by gauge propagators only, then this supergraph cannot contribute to $\DD$ \cite{Sieg:2010tz}.  For example, the supergraph
\begin{equation}
\begin{aligned}\label{ZY}
\settoheight{\eqoff}{$\times$}%
\setlength{\eqoff}{0.5\eqoff}%
\addtolength{\eqoff}{-11\unitlength}%
\raisebox{\eqoff}{%
\fmfframe(-1,1)(-11,1){%
\begin{fmfchar*}(20,20)
\nomix
\fmfi{photon}{vrm1--vrm5}
\fmfi{photon}{vrm3--vrm5}
\fmfi{photon}{vrm3--vlm1}
\fmfi{photon}{vlm2--vlm5}
\fmfi{photon}{vlm3--vlm5}
\fmfi{photon}{vlm2--vlm4}
\end{fmfchar*}}}
\end{aligned}
\end{equation}
only has connections between neighboring legs via gauge propagators, hence power counting will indicate that this supergraph is finite.  The diagram could have subdivergences, but these will  cancel with   with the subdivergences coming from other diagrams because of the superconformal symmetry.  If all legs in a composite are unconnected to their neighboring legs then power counting does not rule out divergences.   
However, the contribution of each leg comes strictly from the chiral field self-energy, whose divergences  cancel by superconformal invariance.  
Notice that these arguments do not actually rely on the $\NN=4$ supersymmetry.  As long as one has an $\NN=1$ superconformal gauge theory with a well-defined perturbative expansion they are still  valid.   

\section{A conjecture for chiral supergraphs in $\NN=4$ SYM}

In this section we present a conjecture for a class of supergraphs in $\NN=4$ SYM.  In the next section these results will be applied to the cubic Leigh-Strassler model.

To this end, let us consider an operator of length $L$ in the $SU(2)$ sector of $\NN=4$ SYM, where $L>>1$ so that wrapping effects can be ignored.  Let us assume that we have one magnon on top of the chiral primary groundstate with momentum $p$.  Strictly speaking this will not be a physical operator, but  we can assume that there is another magnon  on the spin-chain with opposite momentum, and  if we further assume that the magnon is far away then it has  a minimal effect on the other magnon's energy.  The magnon can be represented by the state
\be
|p\rangle\equiv\sum_{j=1}^L \stackrel{\ \ \ \ \stackrel{j}{\downarrow}}{e^{ipj}|XXX\dots XYX\dots XXX\rangle}\,,
\ee
 which has an energy
\be
\veps(p)=\sqrt{1+16g^2\sin^2\frac{p}{2}}-1\,.
\ee
We next observe that we can write the identity
\be\label{epcf}
\veps(p)|p\rangle&=&\left(\sqrt{1+4g^2\chi(1)}-1\right)\, |p\rangle\nn\\
&=&\sum_{m=1}^\infty\frac{\Gamma(3/2)}{\Gamma(m+1)\Gamma(3/2-m)}(2g)^{2m}\big[\chi(1)^m\big]_c\,|p\rangle\,,
\ee
where the expression $[\chi(1)^m]_c$ refers to the connected part of the product.  The unconnected chiral functions give zero when acting on $|p\rangle$.  The first several connected products are
\be\label{conex}
[\chi(1)^2]_c&=&2\chi(1)+\chi(1,2)+\chi(2,1)\nn\\
{}[\chi(1)^3]_c&=&4\chi(1)+4\chi(1,2)+4\chi(2,1)+\chi(1,2,3)+\chi(3,2,1)+\chi(1,2,1)+\chi(2,1,2)\nn\\
{}[\chi(1)^4]_c&=&8\chi(1)+12\chi(1,2)+12\chi(2,1)+6\chi(1,2,3)+6\chi(3,2,1)\nn\\
&&\qquad+6\chi(1,2,1)+6\chi(2,1,2)\\
&&\qquad+\chi(1,2,3,4)+\chi(4,3,2,1)+\chi(1,2,3,2)+\chi(2,3,2,1)\nn\\
&&\qquad+\chi(2,1,2,3)+\chi(3,2,1,2)+\chi(1,2,1,2)+\chi(2,1,2,1)\nn
\ee
We can also see the expansion  diagrammatically using the basic $\chi(1)$ building block, \smash{$\settoheight{\eqoff}{$\times$}%
\setlength{\eqoff}{0.5\eqoff}%
\addtolength{\eqoff}{-11\unitlength}%
\raisebox{\eqoff}{%
\fmfframe(-0,2)(-11,8){%
\begin{fmfchar*}(5,5)
\chionen
\end{fmfchar*}}}\ \ .$}
For example, we have
\begin{equation}
\begin{aligned}\label{chimult}
\Bigg[\settoheight{\eqoff}{$\times$}%
\setlength{\eqoff}{0.5\eqoff}%
\addtolength{\eqoff}{-11\unitlength}%
\raisebox{\eqoff}{%
\fmfframe(-1,1)(-11,6){%
\begin{fmfchar*}(10,10)
\chionen
\end{fmfchar*}}}
\quad\times
\settoheight{\eqoff}{$\times$}%
\setlength{\eqoff}{0.5\eqoff}%
\addtolength{\eqoff}{-11\unitlength}%
\raisebox{\eqoff}{%
\fmfframe(-1,1)(-11,6){%
\begin{fmfchar*}(10,10)
\chionen
\end{fmfchar*}}}\  \ \Bigg]_c\ \ =
\settoheight{\eqoff}{$\times$}%
\setlength{\eqoff}{0.5\eqoff}%
\addtolength{\eqoff}{-11\unitlength}%
\raisebox{\eqoff}{%
\fmfframe(-1,1)(-6,3){%
\begin{fmfchar*}(15,15)
\chionetwon
\end{fmfchar*}}}
\ \ +
\settoheight{\eqoff}{$\times$}%
\setlength{\eqoff}{0.5\eqoff}%
\addtolength{\eqoff}{-11\unitlength}%
\raisebox{\eqoff}{%
\fmfframe(-1,1)(-6,3){%
\begin{fmfchar*}(15,15)
\chioneonen
\end{fmfchar*}}}
\ \ +
\settoheight{\eqoff}{$\times$}%
\setlength{\eqoff}{0.5\eqoff}%
\addtolength{\eqoff}{-11\unitlength}%
\raisebox{\eqoff}{%
\fmfframe(-1,1)(-6,3){%
\begin{fmfchar*}(15,15)
\chitwoonen
\end{fmfchar*}}}\ \ =\chi(1,2)+2\chi(1)+\chi(2,1)\,.
\end{aligned}
\end{equation}

After expanding $[\chi(1)^n]_c$ as in (\ref{conex}),  the expression in (\ref{epcf})   can be resummed, where we find
\be
&&\sum_{m=1}^\infty\frac{\Gamma(3/2)}{\Gamma(m+1)\Gamma(3/2-m)}g^{2m}\big[\chi(1)^m\big]_c\nn\\
&&\qquad\qquad=
F_1(g^2)\,\chi(1)+F_2(g^2)\big[\chi(1,2)\!+\!\chi(2,1)\big]\\
&&\qquad\qquad+F_3(g^2)\big[\chi(1,2,3)\!+\!\chi(3,2,1)\!+\!\chi(1,2,1)\!+\!\chi(2,1,2)\big]+\dots\nn
\ee
with  $F_n(g^2)$  given by
\be\label{Fng2}
F_n(g^2)&=&\sum_{m=n}^\infty\frac{\Gamma(3/2)}{m\,\Gamma(3/2-m)\Gamma(n)\Gamma(m-n+1)}\,(2g)^{2m}\nn\\
&=&\frac{(-1)^{n+1}\Gamma(n\!-\!1/2)}{2\sqrt{\pi}\,\Gamma(n\!+\!1)}\,(2g)^{2n}\,{}_2F_1(n\!-\!1/2,n;n\!+\!1;-8g^2)\,.
\ee
Note that the function  $F_n(g^2)$ that appears in front of a connected chiral function $\chi(a_1,\dots a_n)$  depends only on $n$ and not the particular values of the $a_j$ (although we assume that $|a_j-a_{j+1}|=1$).  Even though the function is written in terms of a hypergeometric function for general $n$, all functions are algebraic.  The first three examples of  these functions are
\be
F_1(g^2)&=&\frac{1}{2}(\sqrt{1\!+\!8g^2}-1)\nn\\
F_2(g^2)&=&-\frac{1}{8}\frac{\left(\sqrt{1\!+\!8g^2}\!-\!1\right)^2}{\sqrt{1\!+\!8g^2}}\nn\\
F_3(g^2)&=&\frac{(\sqrt{1\!+\!8g^2}\!-\!1)^3(3\sqrt{1\!+\!8g^2}\!+\!1)}{64(1\!+\!8g^2)^{3/2}}\,.
\ee

There is one caveat concerning this construction.  Many of the chiral functions are equivalent in the $SU(2)$ sector.  In particular, in the SU(2) sector the chiral functions satisfy the following identity
\be\label{SU2id}
\chi(a_1\dots a_j,a_j+1,a_j\dots a_n)=\chi(a_1\dots a_j,a_j-1,a_j\dots a_n)=\chi(a_1\dots a_j\dots a_n)\,.
\ee
Since
 a single magnon can always be placed in the $SU(2)$ sector under a global  transformation of the  $SU(2|2)\ltimes SU(2|2)$  symmetry group, it might not be meaningful to break up the energy as in (\ref{epcf}).

Recently, however, Sieg  explicitly computed the $SU(2)$ sector dilatation operator in superspace to three loop order \cite{Sieg:2010tz}.  The three loop prediction was first made in \cite{Beisert:2003tq} based on integrability, where in the language of chiral functions it is given by\footnote{We drop a term that does not affect the spectrum.}
\be\label{D3}
\DD_3=4\big(\chi(1,2,3)\!+\!\chi(3,2,1)\big)-4\chi(1,3)+16\big(\chi(1,2)\!+\!\chi(2,1)\big)+24\chi(1).
\ee
One can see that this is consistent with the dispersion relation in (\ref{epcf}) by using $\chi(1,2,1)=\chi(2,1,2)=\chi(1)$.  In Sieg's computation, he found that the following diagrams contribute to $\DD_3$ with a $\chi(1)$ factor:
\begin{equation}\label{chionegraphs}
\begin{aligned}
2\Bigg(\settoheight{\eqoff}{$\times$}%
\setlength{\eqoff}{0.5\eqoff}%
\addtolength{\eqoff}{-12\unitlength}%
\raisebox{\eqoff}{%
\fmfframe(-0.5,2)(-5.5,2){%
\begin{fmfchar*}(20,20)
\chioneg
\fmfi{photon}{vu3--vgm3}
\fmfi{photon}{vd3--vgm3}
\end{fmfchar*}}}
\ +\ 
\settoheight{\eqoff}{$\times$}%
\setlength{\eqoff}{0.5\eqoff}%
\addtolength{\eqoff}{-12\unitlength}%
\raisebox{\eqoff}{%
\fmfframe(-0.5,2)(-5.5,2){%
\begin{fmfchar*}(20,20)
\chioneg
\fmfi{photon}{vm3--vg2}
\fmfi{photon}{vm5--vg2}
\end{fmfchar*}}}
\ + \ 
\settoheight{\eqoff}{$\times$}%
\setlength{\eqoff}{0.5\eqoff}%
\addtolength{\eqoff}{-12\unitlength}%
\raisebox{\eqoff}{%
\fmfframe(-0.5,2)(-5.5,2){%
\begin{fmfchar*}(20,20)
\chioneg
\fmfi{photon}{vu5--vgm5}
\fmfi{photon}{vd5--vgm5}
\end{fmfchar*}}}
\ + \ 
\settoheight{\eqoff}{$\times$}%
\setlength{\eqoff}{0.5\eqoff}%
\addtolength{\eqoff}{-12\unitlength}%
\raisebox{\eqoff}{%
\fmfframe(-0.5,2)(-10.5,2){%
\begin{fmfchar*}(20,20)
\chione
\fmfcmd{fill fullcircle scaled 8 shifted vm4 withcolor black ;}
\fmfiv{label=$\scriptstyle\textcolor{white}{2}$,l.dist=0}{vm4}
\end{fmfchar*}}}
\Bigg)\ +\ 
\settoheight{\eqoff}{$\times$}%
\setlength{\eqoff}{0.5\eqoff}%
\addtolength{\eqoff}{-12\unitlength}%
\raisebox{\eqoff}{%
\fmfframe(-0.5,2)(-10.5,2){%
\begin{fmfchar*}(20,20)
\chione
\fmfcmd{fill fullcircle scaled 8 shifted vloc(__vc2) withcolor black ;}
\fmfv{label=$\scriptstyle\textcolor{white}{2}$,l.dist=0}{vc2}
\end{fmfchar*}}}
\end{aligned}
\end{equation}
where the circled 2 indicates a two-loop self-energy or vertex correction.  Both the self-energy and the vertex correction must be finite by the superconformal invariance.  The overall factor of $2$ in front of the first four diagrams takes into account their reflections.  Note that all of these supergraphs have two neighboring legs connected by a chiral vertex, but if there is a  third leg it is only connected by vector propagators.    The resulting contribution to $\DD_3$ from these supergraphs is \cite{Sieg:2010tz}
\be
\Delta^{(1)} \DD_3=16\chi(1)\,.
\ee

There are two further supergraphs that contribute to $\chi(1)$, which do so through a $\chi(1,2,1)$ or a $\chi(2,1,2)$ structure.  These are 
\begin{equation}\label{chiotogr}
\begin{aligned}
\settoheight{\eqoff}{$\times$}%
\setlength{\eqoff}{0.5\eqoff}%
\addtolength{\eqoff}{-12\unitlength}%
\raisebox{\eqoff}{%
\fmfframe(-0.5,2)(-5.5,2){%
\begin{fmfchar*}(20,20)
\chionetwoonen
\end{fmfchar*}}}
\ +\ 
\settoheight{\eqoff}{$\times$}%
\setlength{\eqoff}{0.5\eqoff}%
\addtolength{\eqoff}{-12\unitlength}%
\raisebox{\eqoff}{%
\fmfframe(-0.5,2)(-5.5,2){%
\begin{fmfchar*}(20,20)
\chitwoonetwon
\end{fmfchar*}}}
\end{aligned}
\end{equation}
which give the contribution  \cite{Sieg:2010tz}
\be
\Delta^{(1,2,1)} \DD_3+\Delta^{(2,1,2)} \DD_3=4\chi(1,2,1)+4\chi(2,1,2)\,.
\ee
Hence,  even though $\chi(1,2,1)$ and $\chi(2,1,2)$ are identified with $\chi(1)$ in the $SU(2)$ sector, the linear combination of connected chiral functions from the three loop dilatation operator is the same linear combination as in  $[\chi(1)^3]_c$.

Base on this evidence, t is tempting to conjecture 
that the all-loop sum of supergraphs with the connected chiral function $\chi(a_1,a_2\dots a_j,\dots a_n)$ is given by
\be\label{F_n}
F_n(g^2)\chi(a_1,a_2\dots a_j,\dots a_n)\,,
\ee
where $F_n(g^2)$ is defined in (\ref{Fng2}).
Note that the sum would include the contributions of all possible insertions of gauge vertices and propagators as well as   chiral loops.  
However, studies at four loops and higher show that this is not the case \cite{MinSieg}.  Instead we make the weaker statement that (\ref{F_n}) captures the rational contribution to the chiral functions.  Starting at four loops there are higher transcendental terms that contribute to the chiral functions, but cancel out in the $\NN=4$ dispersion relation because of the identity in (\ref{SU2id}).
We can still use the $\NN=4$ dispersion and the weaker conjecture to make statements about higher loop contributions to the anomalous dimensions in $\NN=1$ models.


\section{Application to the cubic Leigh-Strassler  model}
\subsection{The cubic model and its corrections to the coupling}
In this section we apply the results from the previous two sections to the cubic Leigh-Strassler model. 

Let us quickly review where the identification $\hh=\sqrt{2}\,g_\YM$ comes from.  In order for the theory to be superconformal, the anomalous dimensions of the chiral fields must be zero.  In superspace, this means that the self-energy for the chiral fields is finite.  If we had chosen the superpotential in (\ref{superpot}) to have $h=0$, $q=1$ and $\kappa=g_\YM$, then this would be the superpotential for $\NN=4$ SYM where we know that all self-energies are finite.  Hence, it is just a question of comparing the difference using the different chiral vertices.  At the one-loop level the relevant supergraph for the self-energy of, say, a $Z$ chiral field is the skeleton\footnote{We will call any supergraph with only chiral propagators a skeleton.
}
\begin{equation}
\begin{aligned}\label{1lose}
Z\ \swfoneone\ \bar Z
\end{aligned}
\end{equation}
In the case of planar $\NN=4$ SYM there are two ways to assign the $X$ and $Y$ flavors on the internal propagators, but for the superpotential in (\ref{Z3superpot}) there is only one way.  Hence, to compensate for this difference, $\hh$ should be chosen  equal $\sqrt{2}\,g_\YM$ so that the contributions of this supergraph to the self-energy are the same.  The divergent part from this supergraph will then cancel with the one-loop supergraphs having only gauge vertices.

Remarkably, this counting works all the way up to the four loop-level, where one encounters a new type of graph that requires a modification of the $\hh$ value of order $\Delta\hh\sim g_\YM^7N^3$ \cite{Bork:2007bj}.  At four loops one has the self-energy skeleton \cite{Elmetti:2006gr,Elmetti:2007up}
\begin{equation}
\begin{aligned}\label{4lose}
Z\ \swffourone\ \bar Z
\end{aligned}
\end{equation}
which has four chiral vertices and 4 anti-chiral vertices.  In the case of planar $\NN=4$ SYM one finds that there are 8 ways to assign the flavors on the internal propagators.  However, since $\hh$ has been set to $\sqrt{2}\,g_{YM}$, this supergraph will come with a factor of 16 for the cubic model.  Hence, there is no longer a matching.  To compensate, one must adjust $\hh$ to $\sqrt{2}g_\YM+\Delta\hh$ so that the scalar field anomalous dimension remains zero to order $g^8$.   Diagrammatically the correction looks like
\begin{equation}
\begin{aligned}\label{1losecorr}
Z\ \settoheight{\eqoff}{$\times$}%
\setlength{\eqoff}{0.5\eqoff}%
\addtolength{\eqoff}{-7.5\unitlength}%
\raisebox{\eqoff}{%
\fmfframe(1,0)(1,0){%
\begin{fmfchar*}(20,15)
\swfoneonel
\fmfcmd{fill fullcircle scaled 8 shifted vloc(__vc1) withcolor black ;}
\fmfv{label=$\scriptstyle\textcolor{white}{\Delta}$,l.dist=0}{vc1}
\end{fmfchar*}}}
\ \bar Z\ +\ 
Z\ \settoheight{\eqoff}{$\times$}%
\setlength{\eqoff}{0.5\eqoff}%
\addtolength{\eqoff}{-7.5\unitlength}%
\raisebox{\eqoff}{%
\fmfframe(1,0)(1,0){%
\begin{fmfchar*}(20,15)
\swfoneonel
\fmfcmd{fill fullcircle scaled 8 shifted vloc(__vc2) withcolor black ;}
\fmfv{label=$\scriptstyle\textcolor{white}{\Delta}$,l.dist=0}{vc2}
\end{fmfchar*}}}
\ \bar Z
\end{aligned}
\end{equation}
where the blob with the $\Delta$ denotes the tree-level correction to the chiral vertex coming from  the correction to $\hh$.  

As was  pointed out in \cite{Elmetti:2006gr,Elmetti:2007up}, canceling the contributions to the anomalous dimension is not the same as canceling the divergence in the chiral propagator since the contributions come at different loop orders.  Using dimensional reduction, the divergence coming from the graphs in (\ref{1losecorr}) and the  extra contribution in (\ref{4lose}) equals  
\be\label{div}
\frac{A\,g^2\,\delta\hh}{g_\YM\,\eps}\left(\frac{\mu^2}{p^2}\right)^{\eps}+\frac{B\,g^8}{\eps}\left(\frac{\mu^2}{p^2}\right)^{4\eps}\,,
\ee
where $A$ and $B$ are some constants, whose actual values can be found in \cite{Bork:2007bj}, but  are not necessary for us here.    Hence, to cancel the anomalous dimension, one should set \be\label{cancad}
A\,\delta\hh+4\,g_\YM\,g^6\,B=0\,. 
\ee
 This leaves the divergence in (\ref{div}) uncanceled.   

There are two proposed ways to deal with this mismatch of divergences, which turn out to be equivalent, at least for everything computed in the latter part of this section.  One way is to include $\epsilon$ dependent corrections in $\hh$ with lower powers of $g_\YM$ \cite{Rossi:2006mu,Kazakov:1986bs,Bork:2007bj,Kazakov:2007dy}.  In this case the $\eps^{-1}$ poles can be canceled with $\eps^{-4}$ poles and lower.   In this sense the theory is finite.   

The other option is to include infinite counterterms for  the chiral propagators and gauge vertices \cite{Elmetti:2007up}, which are related by the Slavnov-Taylor identities.   We will use this procedure here because the bookkeeping strikes us as simpler.  Hence, at the four loop level we would add the counterterm $-(A\,\hh+B\,g^6){\eps^{-1}}$ to the chiral propagator to cancel the divergence.   At five-loops not only will there be new single poles, but there will be double poles as well that need to be canceled.  One can keep going to higher and higher powers of $g^2$ in a systematic fashion, canceling the anomalous dimension and the poles, as well as contributions coming from more complicated skeletons, by adding corrections to $\hh$ and  counterterms.  This process does introduce scheme dependence \cite{Elmetti:2007up} because the counterterms can include finite pieces which affect the next order anomalous dimension cancelation.  Hence the five-loop correction to  $\hh$ is scheme dependent.  However, in the five-loop computation we present later the result is scheme independent, assuming that one uses the same regularization scheme for the operators as for the chiral fields.  

\subsection{Planar protected states}

In the scalar sector, the one-loop Hamiltonian of the cubic model is that of an $SU(3)$ generalization of the anti-ferromagnetic Ising model.   Its explicit form is
\be\label{cubicchione}
g^2\,\DD_1=4\,g^2\,\sum_{j=1}^L\left(\textstyle{\frac{1}{3}}+2\,\tau^3_j\tau^3_{j+1}+\textstyle{\frac{3}{2}}\,\tau^8_j\tau^8_{j+1}\right)\equiv2\,g^2\,\chi(1)\,,
\ee
where $\tau^3$ and $\tau^8$ are the $SU(3)$ generators 
\be
\tau^3=\diag\left(\sfrac{1}{2},-\sfrac{1}{2},0\right)\,,\qquad\tau^8=\diag\left(\sfrac{1}{3},\sfrac{1}{3},-\sfrac{2}{3}\right)\,.
\ee
This defines the chiral function $\chi(1)$ for the cubic model.
  The spin-chain has no kinetic term and one can easily verify that the contribution from neighboring sites with the same flavors is $4g^2$, while  the contribution from neighboring sites with different flavors is $0$.  

In order to investigate the higher loops,  we can borrow many of the arguments from the preceding sections since the coupling in the cubic model is assumed to be tuned so that  there is $\NN=1$  superconformal symmetry.   For example, if no nearest neighbors in a supergraph are connected by a chiral vertex, then the supergraph is finite.  The corrections to $\hh$ and the counterterms do not change this argument because the subdivergences will continue to cancel.  Given the superpotential in (\ref{Z3superpot}), this implies that if all  nearest neighbors in the scalar composite  have different flavors then all supergraphs are finite at the planar level since there is no way to connect neighboring fields with a chiral vertex.   Phrased differently, all chiral functions for the cubic model acting on such an operator are zero.  Hence, in the planar limit, all such operators have zero anomalous dimension for all values of the coupling.   

We will refer to these operators as ``planar protected".  They are not chiral primaries, because  nonplanar supergraphs  can contribute to their anomalous dimensions.  This is also evident from the symmetry, which does not have individual conserved charges for the different  flavors.  For  scalar composites of length $L$, $L>>1$, there are $\sim 2^L/L$ different planar protected composites.     While perhaps obvious to many, we explain the counting in the appendix.

\subsection{The one-pair states}
We next consider  scalar composites where one pair of nearest neighbors have the same flavor but all other nearest neighbors have different flavors.    We call these one-pair states.  The only possibility for a divergent supergraph is that the nearest neighbors in the pair are connected by a chiral vertex.  Since the outgoing fields are the same as the incoming ones, these composites do not mix at the planar level  to all orders in the coupling.  Hence, the position of the pair is fixed on the composite, up to the overall cyclic symmetry.  

For these states,  only the $\chi(1)$  supergraphs contribute.  Hence, up to three loop level we can borrow the $\chi(1)$ contribution of the $\NN=4$ dilatation operator, replacing the $\chi(1)$ with the one found in  (\ref{cubicchione}) to compute the anomalous dimension of this operator.  Taking the expansion of $F_1(g^2)$ in (\ref{Fng2}) and using that $\chi(1)=2$ for the one-pair states gives
\be
\delta^{(3)}_{XX}=4g^2-8g^4+16 g^6+{\rm O}(g^8)\,,
\ee
where the subscript indicates that this is for a one-pair state.
We can  use the $\NN=4$ dispersion relation to find the four-loop contribution.  If we use the conjecture from the previous section we can also find the rational contributions to $\delta_{XX}$ beyond four loops.    

However, since there is a four-loop correction to $\hh$ it is first necessary to adjust the contributing supergraphs.  
At four loops one has the supergraph
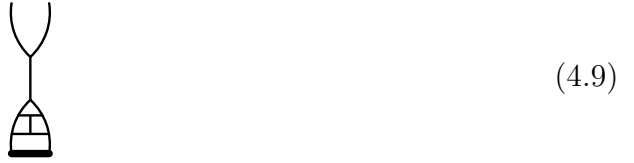
\begin{equation}
\begin{aligned}\label{Z4}
\settoheight{\eqoff}{$\times$}%
\setlength{\eqoff}{0.5\eqoff}%
\addtolength{\eqoff}{-11\unitlength}%
\raisebox{\eqoff}{%
\fmfframe(-1,1)(-11,1){%
\begin{fmfchar*}(20,20)
\chionefour
\end{fmfchar*}}}
\end{aligned}
\end{equation}
One can easily check that this supergraph has an extra factor of 2 in the cubic model as compared to the $\NN=4$ case, just like the self-energy.
To this we should add the correction
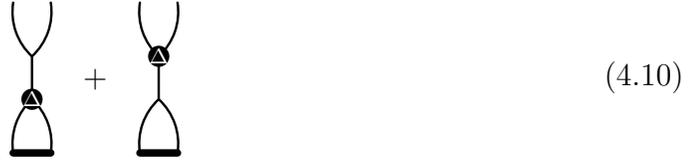
\begin{equation}
\begin{aligned}\label{Z4corr}
\settoheight{\eqoff}{$\times$}%
\setlength{\eqoff}{0.5\eqoff}%
\addtolength{\eqoff}{-12\unitlength}%
\raisebox{\eqoff}{%
\fmfframe(-0.5,2)(-10.5,2){%
\begin{fmfchar*}(20,20)
\chione
\fmfcmd{fill fullcircle scaled 8 shifted vloc(__vc2) withcolor black ;}
\fmfv{label=$\scriptstyle\textcolor{white}{\Delta}$,l.dist=0}{vc2}
\end{fmfchar*}}}
\ + \ 
\settoheight{\eqoff}{$\times$}%
\setlength{\eqoff}{0.5\eqoff}%
\addtolength{\eqoff}{-12\unitlength}%
\raisebox{\eqoff}{%
\fmfframe(-0.5,2)(-10.5,2){%
\begin{fmfchar*}(20,20)
\chione
\fmfcmd{fill fullcircle scaled 8 shifted vloc(__vc1) withcolor black ;}
\fmfv{label=$\scriptstyle\textcolor{white}{\Delta}$,l.dist=0}{vc1}
\end{fmfchar*}}}
\end{aligned}
\end{equation}
  Since the topologies of the loops and comparative factors in (\ref{Z4}) and (\ref{Z4corr}) are the same as in (\ref{4lose}) and (\ref{1losecorr}) respectively, the divergent part in (\ref{Z4corr}) is exactly the same as in (\ref{div}).  Hence, canceling the anomalous dimension for the chiral fields ensures that there is no correction to the anomalous dimension at order $g^8$.    All other four-loop supergraphs lead to the same result as in $\NN=4$ SYM.   
  
  There is still an uncanceled divergence which should be removed by adding a counterterm for the operator.  To find the appropriate counterterm, we note that the equations of motion set
  \be\label{eom}
  XX=\frac{i}{\hat h}\int d^2\bar\theta e^{-g_\YM V}\bar X e^{g_\YM V}\,.
  \ee
Since the righthand side comes from the $D$ term in the Lagrangian, it will have the same wave-function renormalization factor as the chiral propagator.  Hence, the operator counterterm can be implemented by replacing  $XX$ with the righthand side of (\ref{eom}) and with the same counterterm factor as for the chiral propagator.  Diagramatically, we draw this term as
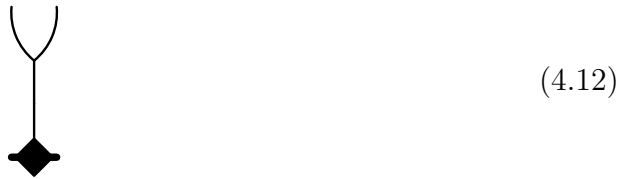
\begin{equation}
\begin{aligned}\label{opct}
\settoheight{\eqoff}{$\times$}%
\setlength{\eqoff}{0.5\eqoff}%
\addtolength{\eqoff}{-12\unitlength}%
\raisebox{\eqoff}{%
\fmfframe(-0.5,2)(-10.5,2){%
\begin{fmfchar*}(20,20)
\ctone
\end{fmfchar*}}}
\end{aligned}
\end{equation}
where one should keep in mind that the upper vertex does not come with a factor of $\hh$ since it has been divided out by the $\hh$ in the denominator of the counterterm.   The divergence in  (\ref{opct}) along with those in (\ref{Z4corr}) now cancel in a way parallel to the cancellation of divergences for the chiral propagator.
  
  We can extend the argument to five loops.  In this case there can be one vector propagator added to (\ref{Z4}) and (\ref{Z4corr}) in various configurations.  For example, one  has the supergraphs
\begin{equation}
\begin{aligned}\label{Z452}
\settoheight{\eqoff}{$\times$}%
\setlength{\eqoff}{0.5\eqoff}%
\addtolength{\eqoff}{-11\unitlength}%
\raisebox{\eqoff}{%
\fmfframe(-1,1)(-11,1){%
\begin{fmfchar*}(20,20)
\chionefour
\fmfi{photon}{vm3{dir 0}..{dir180}vt1}
\end{fmfchar*}}}
\ + \ 
\settoheight{\eqoff}{$\times$}%
\setlength{\eqoff}{0.5\eqoff}%
\addtolength{\eqoff}{-11\unitlength}%
\raisebox{\eqoff}{%
\fmfframe(-1,1)(-11,1){%
\begin{fmfchar*}(20,20)
\chionefour
\fmfi{photon}{vm3{dir 0}..{dir180}vt2}
\end{fmfchar*}}}
\ + \ 
\settoheight{\eqoff}{$\times$}%
\setlength{\eqoff}{0.5\eqoff}%
\addtolength{\eqoff}{-11\unitlength}%
\raisebox{\eqoff}{%
\fmfframe(-1,1)(-11,1){%
\begin{fmfchar*}(20,20)
\chionefour
\fmfi{photon}{vm3{dir 0}..{dir180}vt3}
\end{fmfchar*}}}
\ + \ 
\settoheight{\eqoff}{$\times$}%
\setlength{\eqoff}{0.5\eqoff}%
\addtolength{\eqoff}{-12\unitlength}%
\raisebox{\eqoff}{%
\fmfframe(-0.5,2)(-10.5,2){%
\begin{fmfchar*}(20,20)
\chione
\fmfcmd{fill fullcircle scaled 8 shifted vloc(__vc2) withcolor black ;}
\fmfv{label=$\scriptstyle\textcolor{white}{\Delta}$,l.dist=0}{vc2}
\fmfi{photon}{vm3{dir 0}..{dir180}vm5}
\end{fmfchar*}}}
\ + \ 
\settoheight{\eqoff}{$\times$}%
\setlength{\eqoff}{0.5\eqoff}%
\addtolength{\eqoff}{-12\unitlength}%
\raisebox{\eqoff}{%
\fmfframe(-0.5,2)(-10.5,2){%
\begin{fmfchar*}(20,20)
\chione
\fmfcmd{fill fullcircle scaled 8 shifted vloc(__vc1) withcolor black ;}
\fmfv{label=$\scriptstyle\textcolor{white}{\Delta}$,l.dist=0}{vc1}
\fmfi{photon}{vm3{dir 0}..{dir180}vm5}
\end{fmfchar*}}}
\end{aligned}
\end{equation}
as well as their reflections.
These supergraphs  have the same topologies as the following contributions to the  5 loop self-energy
\begin{equation}
\begin{aligned}\label{4lose51}
\settoheight{\eqoff}{$\times$}%
\setlength{\eqoff}{0.5\eqoff}%
\addtolength{\eqoff}{-7.5\unitlength}%
\raisebox{\eqoff}{%
\fmfframe(1,0)(1,0){%
\begin{fmfchar*}(20,15)
\swffouronel
\fmfi{photon}{vn1{dir 90}..{dir-60}vt1}
\end{fmfchar*}}}
+
\settoheight{\eqoff}{$\times$}%
\setlength{\eqoff}{0.5\eqoff}%
\addtolength{\eqoff}{-7.5\unitlength}%
\raisebox{\eqoff}{%
\fmfframe(1,0)(1,0){%
\begin{fmfchar*}(20,15)
\swffouronel
\fmfi{photon}{vn1{dir 90}..{dir-90}vt2}
\end{fmfchar*}}}
+
\settoheight{\eqoff}{$\times$}%
\setlength{\eqoff}{0.5\eqoff}%
\addtolength{\eqoff}{-7.5\unitlength}%
\raisebox{\eqoff}{%
\fmfframe(1,0)(1,0){%
\begin{fmfchar*}(20,15)
\swffouronel
\fmfi{photon}{vn1{dir 90}..{dir-120}vt3}
\end{fmfchar*}}}
+
\settoheight{\eqoff}{$\times$}%
\setlength{\eqoff}{0.5\eqoff}%
\addtolength{\eqoff}{-7.5\unitlength}%
\raisebox{\eqoff}{%
\fmfframe(1,0)(1,0){%
\begin{fmfchar*}(20,15)
\swfoneonel
\fmfcmd{fill fullcircle scaled 8 shifted vloc(__vc1) withcolor black ;}
\fmfv{label=$\scriptstyle\textcolor{white}{\Delta}$,l.dist=0}{vc1}
\fmfi{photon}{vm3{dir 90}..{dir-90}vm1}
\end{fmfchar*}}}
+
\settoheight{\eqoff}{$\times$}%
\setlength{\eqoff}{0.5\eqoff}%
\addtolength{\eqoff}{-7.5\unitlength}%
\raisebox{\eqoff}{%
\fmfframe(1,0)(1,0){%
\begin{fmfchar*}(20,15)
\swfoneonel
\fmfcmd{fill fullcircle scaled 8 shifted vloc(__vc2) withcolor black ;}
\fmfv{label=$\scriptstyle\textcolor{white}{\Delta}$,l.dist=0}{vc2}
\fmfi{photon}{vm3{dir 90}..{dir-90}vm1}
\end{fmfchar*}}}
\end{aligned}
\end{equation}
where the corresponding diagrams for the reflections have the vector propagators below the external legs.  There are many other supergraphs  similar to (\ref{Z452}), but where both ends of the vector propagator are attached to points below the anti-chiral vertex.  Each one of these supergraphs is equivalent to a particular five-loop self-energy diagram.  Moreover, there is an equivalence for the terms involving counterterms, namely
\begin{equation}
\begin{aligned}\label{opctl}
\settoheight{\eqoff}{$\times$}%
\setlength{\eqoff}{0.5\eqoff}%
\addtolength{\eqoff}{-12\unitlength}%
\raisebox{\eqoff}{%
\fmfframe(-0.5,2)(-10.5,2){%
\begin{fmfchar*}(20,20)
\ctone
\fmfi{photon}{vm3{dir 0}..{dir 210}vm5}
\end{fmfchar*}}}
=
\settoheight{\eqoff}{$\times$}%
\setlength{\eqoff}{0.5\eqoff}%
\addtolength{\eqoff}{-7.5\unitlength}%
\raisebox{\eqoff}{%
\fmfframe(1,0)(1,0){%
\begin{fmfchar*}(20,15)
\ctprop
\fmfi{photon}{vm1{dir 90}..{dir-90}vm3}
\end{fmfchar*}}}
\end{aligned}
\end{equation}

Then there are the five-loop supergraphs involving three legs, for example
\begin{equation}
\begin{aligned}\label{Z453}
\settoheight{\eqoff}{$\times$}%
\setlength{\eqoff}{0.5\eqoff}%
\addtolength{\eqoff}{-11\unitlength}%
\raisebox{\eqoff}{%
\fmfframe(-1,1)(-11,1){%
\begin{fmfchar*}(20,20)
\chionefourg
\fmfi{photon}{vm3--vgm3}
\end{fmfchar*}}}\qquad+
\settoheight{\eqoff}{$\times$}%
\setlength{\eqoff}{0.5\eqoff}%
\addtolength{\eqoff}{-11\unitlength}%
\raisebox{\eqoff}{%
\fmfframe(-1,1)(-11,1){%
\begin{fmfchar*}(20,20)
\chionefourg
\fmfi{photon}{vf55--vgf55}
\end{fmfchar*}}}\qquad+
\settoheight{\eqoff}{$\times$}%
\setlength{\eqoff}{0.5\eqoff}%
\addtolength{\eqoff}{-11\unitlength}%
\raisebox{\eqoff}{%
\fmfframe(-1,1)(-11,1){%
\begin{fmfchar*}(20,20)
\chionefourg
\fmfi{photon}{vm5--vgm5}
\end{fmfchar*}}}\qquad+
\settoheight{\eqoff}{$\times$}%
\setlength{\eqoff}{0.5\eqoff}%
\addtolength{\eqoff}{-11\unitlength}%
\raisebox{\eqoff}{%
\fmfframe(-1,1)(-11,1){%
\begin{fmfchar*}(20,20)
\chionefourg
\fmfi{photon}{vf51--vgf51}
\end{fmfchar*}}}
\end{aligned}
\end{equation}  
as well as their reflections.  Using the $D$-algebra, one can show that the first two and the last two each sum to a UV finite result \cite{Fiamberti:2008sh,Sieg:2010tz}, although this is not  crucial for our argument.  These supergraphs are topologically equivalent to the five-loop self-energy supergraphs
\begin{equation}
\begin{aligned}\label{4lose53}
\settoheight{\eqoff}{$\times$}%
\setlength{\eqoff}{0.5\eqoff}%
\addtolength{\eqoff}{-7.5\unitlength}%
\raisebox{\eqoff}{%
\fmfframe(1,0)(1,0){%
\begin{fmfchar*}(20,15)
\swffouronel
\fmfi{photon}{vn1{dir 90}..{dir-90}vn2}
\end{fmfchar*}}}
+
\settoheight{\eqoff}{$\times$}%
\setlength{\eqoff}{0.5\eqoff}%
\addtolength{\eqoff}{-7.5\unitlength}%
\raisebox{\eqoff}{%
\fmfframe(1,0)(1,0){%
\begin{fmfchar*}(20,15)
\swffouronel
\fmfi{photon}{vt1{dir 120}..{dir-90}vn2}
\end{fmfchar*}}}
+
\settoheight{\eqoff}{$\times$}%
\setlength{\eqoff}{0.5\eqoff}%
\addtolength{\eqoff}{-7.5\unitlength}%
\raisebox{\eqoff}{%
\fmfframe(1,0)(1,0){%
\begin{fmfchar*}(20,15)
\swffouronel
\fmfi{photon}{vt2{dir 90}..{dir-90}vn2}
\end{fmfchar*}}}
+
\settoheight{\eqoff}{$\times$}%
\setlength{\eqoff}{0.5\eqoff}%
\addtolength{\eqoff}{-7.5\unitlength}%
\raisebox{\eqoff}{%
\fmfframe(1,0)(1,0){%
\begin{fmfchar*}(20,15)
\swffouronel
\fmfi{photon}{vt3{dir 60}..{dir-90}vn2}
\end{fmfchar*}}}
\end{aligned}
\end{equation}
If we include the reflections of the supergraphs in   (\ref{4lose53}), then the divergence from all possible five-loop supergraphs involving the basic four-loop skeleton in (\ref{Z4}) equals the divergence from all possible self-energy supergraphs built from the four-loop self-energy skeleton  in (\ref{4lose}).  The same holds true for the two-loop supergraphs built out of (\ref{Z4corr}) and the one-loop term built from the counterterm in (\ref{opct}).  Finally, both the chiral propagator and the operator require the same double-pole counterterm to cancel this divergence.  Hence, if $\hh$ and the counterterms are tuned to cancel  the divergences in the five-loop self-energy, then they also cancel  the  unwanted divergences  of the operator supergraphs.  We emphasize that this is independent of the regularization scheme one chooses.

This matching of operator supergraphs to self-energy supergraphs  can be applied to higher loops as well, as long as the supergraphs involve only two or three legs of the scalar composite.  However,
 starting at six loops one can have supergraphs involving four legs.  The contribution from those built from the skeleton in (\ref{Z4}) are
\begin{equation}
\begin{aligned}
\label{r4chi1cancel}
&
\settoheight{\eqoff}{$\times$}%
\setlength{\eqoff}{0.5\eqoff}%
\addtolength{\eqoff}{-12\unitlength}%
\raisebox{\eqoff}{%
\fmfframe(-0.5,2)(-0.5,2){%
\begin{fmfchar*}(20,20)
\gchionefourg
\fmfi{photon}{vm4--vglm5}
\fmfi{photon}{vm5--vgrm5}
\end{fmfchar*}}}
+
\settoheight{\eqoff}{$\times$}%
\setlength{\eqoff}{0.5\eqoff}%
\addtolength{\eqoff}{-12\unitlength}%
\raisebox{\eqoff}{%
\fmfframe(-0.5,2)(-0.5,2){%
\begin{fmfchar*}(20,20)
\gchionefourg
\fmfi{photon}{vm3--vglm3}
\fmfi{photon}{vm3--vgrm3}
\end{fmfchar*}}}
+
\settoheight{\eqoff}{$\times$}%
\setlength{\eqoff}{0.5\eqoff}%
\addtolength{\eqoff}{-12\unitlength}%
\raisebox{\eqoff}{%
\fmfframe(-0.5,2)(-0.5,2){%
\begin{fmfchar*}(20,20)
\gchionefourg
\fmfi{photon}{vu3--vglu3}
\fmfi{photon}{vd3--vgrd3}
\end{fmfchar*}}}
+
\settoheight{\eqoff}{$\times$}%
\setlength{\eqoff}{0.5\eqoff}%
\addtolength{\eqoff}{-12\unitlength}%
\raisebox{\eqoff}{%
\fmfframe(-0.5,2)(-0.5,2){%
\begin{fmfchar*}(20,20)
\gchionefourg
\fmfi{photon}{vm3--vglm3}
\fmfi{photon}{vfr55--vgfr55}
\end{fmfchar*}}}
+
\settoheight{\eqoff}{$\times$}%
\setlength{\eqoff}{0.5\eqoff}%
\addtolength{\eqoff}{-12\unitlength}%
\raisebox{\eqoff}{%
\fmfframe(-0.5,2)(-0.5,2){%
\begin{fmfchar*}(20,20)
\gchionefourg
\fmfi{photon}{vu3--vglu3}
\fmfi{photon}{vm5--vgrm5}
\end{fmfchar*}}}
+
\settoheight{\eqoff}{$\times$}%
\setlength{\eqoff}{0.5\eqoff}%
\addtolength{\eqoff}{-12\unitlength}%
\raisebox{\eqoff}{%
\fmfframe(-0.5,2)(-0.5,2){%
\begin{fmfchar*}(20,20)
\gchionefourg
\fmfi{photon}{vm3--vglu3}
\fmfi{photon}{vfr51--vgfr51}
\end{fmfchar*}}}
\,,
\end{aligned}
\end{equation}
plus the reflections of the last four.  One can show using the $D$-algebra that the divergences in  these supergraphs cancel in pairs \cite{Fiamberti:2008sh,Sieg:2010tz}.  Furthermore, using the $D$-algebra the last four can be shown equivalent to the six-loop self-energy diagrams
\begin{equation}
\begin{aligned}\label{4lose63}
\settoheight{\eqoff}{$\times$}%
\setlength{\eqoff}{0.5\eqoff}%
\addtolength{\eqoff}{-7.5\unitlength}%
\raisebox{\eqoff}{%
\fmfframe(1,0)(1,0){%
\begin{fmfchar*}(20,15)
\swffouronel
\fmfi{photon}{vnn2{dir 90}..{dir-90}vnn4}
\fmfi{photon}{vnn1{dir -90}..{dir90}vnn3}
\end{fmfchar*}}}
+
\settoheight{\eqoff}{$\times$}%
\setlength{\eqoff}{0.5\eqoff}%
\addtolength{\eqoff}{-7.5\unitlength}%
\raisebox{\eqoff}{%
\fmfframe(1,0)(1,0){%
\begin{fmfchar*}(20,15)
\swffouronel
\fmfi{photon}{vt1{dir 120}..{dir-90}vnn2}
\fmfi{photon}{vnn1{dir -90}..{dir90}vnn3}
\end{fmfchar*}}}
+
\settoheight{\eqoff}{$\times$}%
\setlength{\eqoff}{0.5\eqoff}%
\addtolength{\eqoff}{-7.5\unitlength}%
\raisebox{\eqoff}{%
\fmfframe(1,0)(1,0){%
\begin{fmfchar*}(20,15)
\swffouronel
\fmfi{photon}{vt2{dir 90}..{dir-90}vnn2}
\fmfi{photon}{vnn1{dir -90}..{dir90}vnn3}
\end{fmfchar*}}}
+
\settoheight{\eqoff}{$\times$}%
\setlength{\eqoff}{0.5\eqoff}%
\addtolength{\eqoff}{-7.5\unitlength}%
\raisebox{\eqoff}{%
\fmfframe(1,0)(1,0){%
\begin{fmfchar*}(20,15)
\swffouronel
\fmfi{photon}{vt3{dir 60}..{dir-90}vnn2}
\fmfi{photon}{vnn1{dir -90}..{dir90}vnn3}
\end{fmfchar*}}}
\,.
\end{aligned}
\end{equation}
The first two diagrams in (\ref{r4chi1cancel}) do not have an obvious mapping to  self-energy diagrams, although this is not important since their UV divergent parts cancel.  Likewise, there are six-loop self-energy diagrams that do not have  obvious partners for  operator diagrams.  At this time we have not established whether or not the UV divergences for these unmatched self-energy diagrams cancel because of $D$-algebra identities.

Having established the cancellation of these extra terms to at least five-loop order, we now claim using the conjecture from the previous section 
that the anomalous dimension for the one-pair class of operators  is
 \be\label{fiveloops}
\delta^{(5)}_{X\!X}=2F_1(g^2)+T^{(1)}(g^2)+{\rm O}(g^{12})=\sqrt{1+8g^2}-1+T^{(1)}(g^2)+{\rm O}(g^{12})\,.
\ee
The function $T^{(1)}(g^2)$ contains higher transcendental terms starting at four-loop order.
%

We can find the four-loop contribution to $T^{(1)}(g^2)$ indirectly.  The connected chiral functions that  receive four-loop contributions include $\chi(1)$, $\chi(1,2,1)$ and $\chi(2,1,2)$.  By symmetry, $\chi(1,2,1)$ and $\chi(2,1,2)$ have the same four-loop contribution and both are equivalent to $\chi(1)$ in the $SU(2)$ sector.  If $\chi(1,2,1)$ has a transcendental contribution, since no such term appears in the $\NN=4$ dispersion, it must cancel out with a contribution from $T^{(1)}(g^2)$.   In \cite{MinSieg} we show that the four-loop transcendental contribution from $\chi(1,2,1)$ is $8 \zeta(3)g^8$, hence to cancel this, we must have
\be
T^{(1)}(g^2)=-16\zeta(3)g^8+{\rm O}(g^{10})\,.
\ee
Therefore, the  anomalous dimension, up to and including four-loop order, is
\be
\delta^{(4)}_{XX}=4g^2-8g^4+16 g^6-16(5+\zeta(3))g^8+{\rm O}(g^{10})\,,
\ee
Note that since the four-loop transcendental contribution is negative, we expect that the anomalous dimension scales less than $g$ for large $g$.

Note that result in (\ref{fiveloops}) will be modified by wrapping corrections \cite{Ambjorn:2005wa,Sieg:2005kd,Fiamberti:2007rj,Fiamberti:2008sh,Bajnok:2008bm}.   Just like the case of $\NN=4$ SYM, there is a correction at $L$ loop order for a composite operator of length $L$.

\subsection{Multi-paired states}
We can also consider multiple pairs of fields.  The positions of the pairs are still fixed.  If  pairs are neighbors then they must be composed of different scalar fields.  Even though the pairs are fixed they still have a separation dependent potential.  At weak coupling the potential scales as $g^{2(2+\ell)}$, where $\ell$ is the separation.  This means that the interaction first shows up at three-loop order.   The three-loop dilatation operator in (\ref{D3}) contains the term 
\be
\DD_3^{(1,3)}=-4g^6\chi(1,3)\,.
\ee
Hence, the change in energy for each neighboring pair   is $-16g^6$ to lowest nonzero order.
We conclude that for any scalar composite operator with no more than two sequential flavors of the same type   anywhere in the trace, the anomalous dimension is known to at least three-loop order.  This means that for scalar composites of length $L$, we know the anomalous dimensions for approximately $\frac{1}{L}\left(\frac{11}{4}\right)^L$ different operators, along with their complex conjugates (see the appendix for an explanation of the counting).

\subsection{Quadruplets}
When there are three or more sequential  scalar fields  with the same flavor, the composite operator can mix with other composites and the spin-chain becomes dynamic.  For instance,  $X\!X\!X$ can mix into  $\eps^{\al\beta}W_{\alpha}W_\beta$ through the supergraph
\begin{equation}
\begin{aligned}
\label{3to2}
\settoheight{\eqoff}{$\times$}%
\setlength{\eqoff}{0.5\eqoff}%
\addtolength{\eqoff}{-12\unitlength}%
\raisebox{\eqoff}{%
\fmfframe(-0.5,2)(-0.5,2){%
\begin{fmfchar*}(20,20)
\threetotwo
\end{fmfchar*}}}
\end{aligned}
\end{equation}
where  a $\bar D^2 D_\al$ and a $D_\beta$ are pulled out of the loop and onto the vector legs.  This assumes that  one of the three scalar fields is missing a $\bar D^2$, otherwise we would pull out an extra $\bar D^2$ onto the second vector.

Once produced, the $W_\al$'s can  migrate along the chain, or the pair can mix back into  $X\!X\!X$, $YYY$ or $ZZZ$.  To see what happens in this latter case, suppose  we start with a composite operator that alternates between $X$ and $Y$, except for one sequence of $YYYY$,
\be
\dots XY\!XY\!X[YYYY]\!XY\!XY\!XY\dots
\ee
Then either the left three $Y$'s or the right three can transform into an  $\!X\!X\!X$ through a $W^2$, which is essentially a two-loop effect  \cite{Beisert:2003ys}.   This leaves an $\!X\!X\!X\!X$ either to the left or the right of the original $YYYY$.   The process can then continue.  For example, we could have the pattern
\be
\begin{array}{c}\dots XY\!XY\!X[YYYY]\!XY\!XY\!XY\dots\\
\!\!\downarrow {\scriptstyle W^2}\\
\dots XY\!XY[\!X\!X\!X\!X]YXY\!XY\!XY\dots\\
\!\!\!\!\!\!\!\!\!\!\downarrow {\scriptstyle W^2}\\
\dots XY\!X[YYYY]\!XYXY\!XY\!XY\dots\\
\!\!\!\!\!\!\!\!\!\!\downarrow {\scriptstyle W^2}\\
\dots XY\!XY[\!X\!X\!X\!X]YXY\!XY\!XY\dots
\end{array}
\ee
where we see that the quadruplet of scalar fields can propagate in either direction along the chain.  Its position determines whethere the quadruplet is $\!X\!X\!X\!X$ or $YYYY$.
While  $YYY$ and $X\!X\!X$ can also mix into  $ZZZ$, the $ZZZ$ will be fixed in place and  will eventually transform back into $\!X\!X\!X$ or $YYY$.  Since the triplet $ZZZ$ has less energy than the quadruplet at the one-loop level, where the scalar composite  is nondynamical, this will be a higher order effect.

If we had started with a scalar composite  with only the  triplet  $YYY$, then it could behave as follows
\be
\begin{array}{c}\dots XY\!XY\!X[YYY]\!XY\!XY\!XY\dots\\
\ \ \downarrow {\scriptstyle W^2}\\
\dots XY\!XY[\!XX\!X\!XX]Y\!XY\!XY\dots\\
\!\!\!\!\!\!\!\downarrow {\scriptstyle W^2}\\
\dots XY\!X[YYYY][\!X\!X]Y\!XY\!XY\dots\\
\!\!\!\!\!\!\!\!\!\!\!\!\!\!\!\!\!\downarrow {\scriptstyle W^2}\\
\dots XY[\!X\!X\!X\!X]Y[\!X\!X]Y\!XY\!XY\dots
\end{array}
\ee
Here we see that the triplet has split into a pair and a quadruplet.  The pair stays fixed while the quadruplet migrates along the chain.  The pair then acts as a fixed impurity which the quadruplet can scatter through.  

One can continue to add quadruplets and pairs onto the chain.  It should be straightforward to derive the Hamiltonian for this system up to two-loop order using the dilation operator of the $SU(2|3)$ sector in \cite{Beisert:2003ys}, although we have not done this here.

\subsection{A finite Hagedorn temperature at strong coupling?}
The exponential growth in $L$ of the planar protected states would indicate that the gravity dual of this theory is very unusual.  The gravity dual for the superpotential in (\ref{superpot}) with $q=1$ and $|h|<<1$ has been investigated in \cite{Aharony:2002hx,Kulaxizi:2006zc}.   For the dual one turns on fluxes and deforms the $S^5$.  The cubic model is far away from this regime.  In any case, even at large coupling the exponential growth of the planar protected operators would seem to indicate that the theory on $S^1\times S^3$ has a Hagedorn temperature of order one in units of the $S^3$ inverse radius, even at strong coupling.   At zero coupling the particle content is the  same as $\NN=4$ SYM and hence would have the same Hagedorn temperature \cite{Sundborg:1999ue,Aharony:2003sx}.  Going to infinite coupling in $\NN=4$ SYM  takes the Hagedorn temperature to infinity, since the growth of chiral primaries is  polynomial in $L$.  Of course this is   expected since the  dual theory reduces to supergravity.  For the cubic model, since there appears to be a finite Hagedorn temperature it would indicate that the low energy gravity dual is stringy in nature.

Even though the cubic model appears to have a Hagedorn temperature, it is far from obvious that there would be a sharp phase transition here.  At weak coupling, the Hagedorn temperature indicates a breakdown in planarity.  But if planarity is breaking down, then the nonplanar contributions to the anomalous dimensions  become important.  This could then tamp down the exponential growth in the density of states and push the transition point upward, perhaps all the way to $(\la)^{1/4}$ in the strong coupling limit.

\section{Discussion}

In this paper we 
considered the cubic branch of the Leigh-Strassler superconformal theories, arguing that it has an exponentially large number of  states with zero anomalous dimension in the planar limit.  We then considered states where impurities were added to these zero energy states.  Using 
 information about the $\NN=4$ magnon dispersion relation we were able to compute the anomalous dimension of these operators to 4-loop order.  Furthermore, using a conjecture relating sums of connected graphs we are able to make an all orders prediction for the rational parts of their anomalous dimensions, assuming that unwanted divergences will continue to cancel beyond five-loop order.

There are several directions that one can pursue.  Since part of our results depends on the conjecture, 
it would be very helpful to have a proof (or disproof) of its validity. 
Further evidence for the conjecture will be presented in \cite{MinSieg}.  It would also be desirable to have a better understanding of how the transcendental parts enter into the $\NN=1$ computations.  Our conjecture only uses the dispersion relation and not the $S$-matrix, so perhaps one could find similar relations between the $S$-matrix and classes of supergraphs that are  associated with more general  chiral functions and not just the connected ones.  
It would also be useful to find an argument showing that  the unwanted divergences continue to cancel  beyond five-loop order when adjusting the coupling to cancel the anomalous dimension of the chiral fields.

Another possible direction is to find an analog of the conjecture for the ABJM/ABJ models \cite{Aharony:2008ug,Aharony:2008gk}.  In particular, this could be useful in ascertaining the higher loop corrections to the undetermined function $h^2(\la,\hat{\la})$ that appears in the magnon dispersion relation  \cite{Minahan:2009aq,Minahan:2009wg,Leoni:2010tb} and the $S$-matrix.
Inspecting the supergraphs in  \cite{Leoni:2010tb}, there is some indication that they are organized in a way related to the dispersion relation and the two  't~Hooft couplings.  The spin-chains in  ABJM/ABJ are of alternating type, which leads to two types of magnons.  The dispersion relation is then investigated in the $SU(2)\times SU(2)$ sector.  The first relevant equivalence for chiral functions is $\chi(1,3,1)=\chi(3,1,3)=\chi(1)$, and one would have to do a six-loop calculation to find the equivalent data point of the three-loop calculation in $\NN=4$.  As a very simple check, if the conjecture is true then the contribution from the $\chi(1,3,1)$ and $\chi(3,1,3)$ supergraphs could only be proportional to $\la^3\hat{\la}^3$ based on the leading expansion of $h^2(\la,\hat{\la})$.  One can quickly verify by drawing the relevant diagrams with the $\chi(1,3,1)$ and $\chi(3,1,3)$ chiral functions that this is the case.  

The ABJM/ABJ model is  complicated by the fact that the chiral vertices are four-point and so the supergraphs do not have the multiplication structure of the  fundamental building blocks as in  (\ref{chimult}).  A model that is similar to ABJM/ABJ in that it also has two couplings is the  interpolating $\NN=2$ superconformal $SU(N)\times  SU(N)$ quiver theory discussed in \cite{Gadde:2009dj,Gadde:2010zi}.   When the couplings are equal then this is the $Z_2$ orbifold of $\NN=4$ SYM.  As one varies the ratio of couplings it interpolates between this point and $\NN=2$ SYM with $2N$ hypermultiplets.   The orbifold point has exactly the same supergraph structure as the unorbifolded theory and one can thus borrow all of the results from there.  Moving away from the orbifold point, one starts seeing differences at three-loop level \cite{Pomoni:2011jj}.  The chiral and anti-chiral vertices are always trivalent so the chiral structure is the same as $\NN=4$ and any differences can be swept into the chiral functions which would now also depend on the ratio of couplings \cite{Pomoni:2011jj}.  Hence, one can use the conjecture here, and the problem reduces to finding the chiral functions, although they are  are  highly nontrivial and are known only to the lowest loop order \cite{Pomoni:2011jj}.

It would be very interesting if 
the cubic model were to have any integrable structures.  
It is unlikely that the full theory is integrable, but  perhaps there is integrability in the closed sector of  
chiral composite operators.  
It would also 
be interesting to find the gravity dual for the cubic model.  As stated in the main text, we expect it to have some stringy features, even at low energies.  In essence, there can be no separation into a supergravity 
limit with a finite number of fields.
%

\subsection*{Acknowledgments}
I thank Christoph Sieg for a careful reading of the manuscript and his many helpful comments.   I  also thank the
CTP at MIT   for kind
hospitality  during the course of this work. This  research is supported in part by
Vetenskapr\aa det.
\vfill\eject

\appendix
\section{Appendix}

Here we explain the counting of states for certain types of scalar composites of length $L$.

To count states where all nearest neighbors have different flavors, start with the first site of the chain.  This has 3 choices.  All subsequent sites up to $L\!-\!1$ have 2 choices, since they must be different than the preceding site.  At the last site the flavor must be different than the flavor at the first site and at the $L\!-\!1$ site.  Since we are assuming that $L>>1$ we can assume that for random sequential choices starting from the first site and moving to the right, the first site and the $L\!-\!1$ site are uncorrelated.  Hence, there is a $2/3$ chance that they have different flavors in which case there is only one choice for the last site, and a $1/3$ chance that they have the same flavor, in which case there are 2 choices for the last site.  Thus, the average number of choices for the last site is $4/3$.  Combining the first and the last site we see that on average there are  4 choices for the two of them, leaving $\approx 2^L$ different combinations for the whole chain.
  However, because of the trace, composite operators are identified if  all fields in the operator are shifted  by one site.  This means that for almost all chains of length $L$ (all chains if $L$ is prime), they are identified with $L\!-\!1$ other chains.   Therefore, we should divide $2^L$ by $L$. 

To count scalar composites which do not have three sequential sites with the same flavor, we again assume that $L$ is large.  At any one site on the chain there are 3 choices for the flavor if the preceding two sites have different flavors, and 2 choices if the preceding two sites have the same flavor.  To this end, let $r$ be the probability that the preceding two sites are the same, then the average number of choices for the site are $2r\!+\!3(1\!-\!r)=3\!-\!r$.  In order to be consistent, we must find the same $r$ for the probability that the site and its precedent have the same flavor.  They can only have the same flavor if the preceding two are different, in which case there is a $1/3$ chance.  Therefore, we have the consistency relation $(1\!-\!r)/3=r$.  Hence, $r=1/4$ and the average number of choices is $11/4$.  Again we must divide by $L$ because of the trace condition, giving us $\approx (11/4)^L/L$ different scalar composites of this type.  

Note that the total number of scalar composites with no conditions is $\sim3^L/L$.


\footnotesize
\bibliographystyle{JHEP}
\bibliography{references}

\end{fmffile}
\end{document}